\documentclass[12pt]{amsart}
\theoremstyle{definition} \theoremstyle{remark}
\numberwithin{equation}{section}

\begin{document}
\title[BRST quantization of quasi-symplectic manifolds]{BRST
quantization  of quasi-symplectic manifolds \\ and beyond}
\author{S. L. Lyakhovich}
\address[S. L. Lyakhovich and A. A. Sharapov]{Department of Theoretical Physics,
Physics Faculty, Tomsk State University,   Lenin av. 36, 634050
Tomsk, Russia.} \email[SLL]{sll@phys.tsu.ru}
\author{A. A. Sharapov}
\curraddr[A. A. Sharapov]{} \email[AAS]{sharapov@phys.tsu.ru}

\thanks{The authors are thankful to I.A. Batalin, V.A. Dolgushev and
M.A. Grigoriev for fruitful discussions and to T. Strobl for
useful comments on the manuscript. This research was supported in
part by RFBR under the grants 03-02-17657, 03-02-06709, Russian
Ministry of Education under the grant E 02-3.1-250 and by the
INTAS grant 00-00262. AAS appreciates the financial support of the
Dynasty Foundation and the International Center for Fundamental
Physics in Moscow. SLL is partially supported by the RFBR grant
02-01-00930.}

\begin{abstract}
We consider a class of \textit{factorizable} Poisson brackets
which includes almost all reasonable Poisson structures. A
particular case of the factorizable brackets are those associated
with symplectic Lie algebroids. The BRST theory is applied to
describe the geometry underlying these brackets as well as to
develop a deformation quantization procedure in this particular
case. This can be viewed as an extension of the Fedosov
deformation quantization to a wide class of \textit{irregular}
Poisson structures. In a more general case, the factorizable
Poisson brackets are shown to be closely  connected with the
notion of $n$-algebroid. A simple description is suggested for the
geometry underlying the factorizable Poisson brackets basing on
construction of an odd Poisson algebra bundle equipped with an
abelian connection. It is shown that the zero-curvature condition
for this connection generates all the structure relations for the
$n$-algebroid as well as a generalization of the Yang-Baxter
equation for the symplectic structure.
\end{abstract}

\maketitle

\section{Introduction}

The deformation quantization of a Poisson manifold
$(M,\{\cdot,\cdot\})$ is the construction of a local one-parameter
deformation of the commutative algebra of functions $C^\infty(M)$
respecting associativity \cite{Berezin}, \cite{BFFLS}. The
deformed product is usually denoted by $\ast$, and the deformation
parameter is the Plank constant $\hbar$. In each order in $\hbar$
the $\ast$-product is given by a bi-differential operator
(locality) and the skew-symmetric part of the first $\hbar$-order
coincides with the Poisson bracket of functions (correspondence
principle).

Very early it appeared that the complexity of the deformation
quantization program essentially depends on whether a given
Poisson manifold is regular or not. In the regular case, i.e.,
where the rank of the Poisson tensor is constant, one can
introduce an affine symmetric connection respecting the Poisson
structure (a Poisson connection). Clearly, in the irregular case
such a connection cannot exist. The relevance of the Poisson
connection for constructing $\ast$-products had been already
discussed in \cite{BFFLS}, but in its full strength, the
connection was first exploited by Fedosov in his seminal paper
\cite{Fedosov} on the deformation quantization of symplectic and
regular Poisson manifolds (see also \cite{Fedosovbook}).

The  existence of deformation quantization for general Poisson
manifolds, not necessarily regular, was proved by Kontsevich
\cite{Kontsevich} as a consequence of his Formality Theorem. An
explicitly covariant version of the Kontsevich quantization has
been given in \cite{CFT} (see also \cite{Dolg}, where both
covariant and equivariant versions of the formality theorem have
been presented). It should be noted that the Kontsevich
quantization is based on completely different ideas and involves
more complicated algebraic  technique as compared to the Fedosov
quantization. A nice ``physical explanation" of the Kontsevich
quantization formula was given in \cite{CF} by applying the BV
quantization method \cite{BV1} to the Poisson sigma-model.

Recently, it was recognized that the method of Fedosov's
quantization can further  be extended to include a certain class
of irregular Poisson manifolds even though no Poisson connection
can exist in this case. To give an idea about the manifolds in
question let us write the following expression describing the
general structure of the corresponding Poisson brackets:
\begin{equation}
\{f , g\}= \omega^{ab}(X_a^\mu\partial_\mu f) (X_b^\nu\partial_\nu
g)\, , \qquad \det (\omega^{ab}) \neq 0\,. \label{factor}
\end{equation}
The matrices $X$ and $\omega$ are subject to  certain conditions
ensuring the Jacobi identity.  The geometric meaning of these
conditions as well as the precise mathematical status of $X$ and
$\omega$ will be explained in the next section. Here we would like
to mention that, no a priori assumption is made about the rank of
the matrix $X$, so the Poisson brackets (\ref{factor}) may well be
irregular.

In the case where the matrix $X$ is the anchor of a Lie algebroid
the manifolds under consideration are something intermediate
between symplectic and general Poisson manifolds. For this reason,
we refer to them as \textit{quasi-symplectic} Poisson manifolds
(not to be confused with the quasi-Poisson manifolds introduced in
Ref. \cite{AK}). Being closely related with the notion of a
dynamical $r$-matrix, these manifolds may be of immediate interest
in the theory of integrable systems.

The generalization of the Fedosov deformation quantization to the
case of symplectic Lie algebroids was first given by Nest and
Tsygan \cite{NT}. They also proved corresponding classification
theorems. Fedosov's quantization method was also described in the
work \cite{Vaisman} for the same class of manifolds in the
language of symplectic ringed spaces. Particular classes of
quasi-symplectic manifolds have been quantized in \cite{Rieffel},
\cite{Xu}, \cite{DILSh} making use of various ideas, including
BRST theory.

The aim of this work is twofold. In the first part of the paper we
put  the deformation quantization of quasi-symplectic manifolds in
the framework of BFV-BRST theory \cite{BFV}, \cite{BFV2},
\cite{HT}. For the (constrained) Hamiltonian systems on symplectic
manifolds, the relationship has been already established  between
the BFV-BRST and the Fedosov quantizations \cite{GL},\cite{BGL}.
Here we re-shape this technology to make it working in a more
general case of quasi-symplectic manifolds. The second part of the
paper is devoted to a possible generalization of the notion of a
quasi-symplectic manifold to the case of $n$-algebroids or, in
other terminology,  NQ-manifolds \cite{Vain}, \cite{Sev},
\cite{Vor}, \cite{Roy}. This generalization essentially relaxes
the restrictions on the structure functions $X$ and $\omega$,
entering factorization (\ref{factor}), and covers almost all
reasonable Poisson structures.

The paper is organized as follows. In Sect.2 we give the
definition of a quasi-symplectic Poisson manifold  and discuss
some examples. Here we also construct a simple counter-example to
existence of a quasi-symplectic representation for any Poisson
bracket. Sect.3 deals with realization of quasi-symplectic
manifolds as coisotropic surfaces in the total space of vector
bundles associated with symplectic Lie algebroids. In Sect.4 this
realization is exploited to perform the BRST quantization of the
resulting gauge system. We prove that the quantum multiplication
in the algebra of physical observables induces an associative
$\ast$-product on the initial quasi-symplectic manifold. In Sect.5
we generalize the notion of a quasi-symplectic manifold to a wider
class of factorizable Poisson brackets.  Under reasonable
restrictions this class of Poisson structures is proved to be
closely connected  with $n$-algebroids. Using the 2-algebroid as
example, we show how the geometry underlying factorizable Poisson
brackets can be described in terms of a super-vector bundle
equipped with a fiber-wise odd Poisson structure and a compatible
abelian connection.

\section{Quasi-symplectic manifolds: definition and examples}

The most concise and geometrically transparent way to define the
quasi-symplectic manifolds is to use the notion of a \textit{Lie
algebroid\/} \cite{WCdS}.

\vspace{5mm} \noindent {\bf Definition.} A Lie algebroid over a
manifold $M$ is a (real) vector bundle $\mathcal{E}\rightarrow M$
equipped with the following additional structures.
\begin{enumerate}
    \item There is a (real) Lie algebra structure on the linear space
     of sections $\Gamma(\mathcal{E})$.
     \item There is a bundle map $\rho:
\mathcal{E}\rightarrow TM$ such that the Lie algebra and
$C^{\infty}(M)$-module structures on $\Gamma(\mathcal{E})$ are
compatible in the following sense:
\begin{equation}\label{anc}
[s_1,fs_2]=f[s_1,s_2]+(\rho_\ast(s_1)f)s_2 \,, \qquad \forall f\in
C^{\infty}(M)\,,\quad \forall s_1,s_2\in \Gamma(\mathcal{E})\,.
\end{equation}
\end{enumerate}
The map $\rho$ is called the \textit{anchor} of the Lie algebroid
$\mathcal{E}\rightarrow M$. \vspace{5mm}

The last relation can be viewed as the Leibniz rule for the Lie
algebroid bracket. Using this relation and the Jacobi identity for
the bracket it is not hard to see that the anchor map $\rho:
\mathcal{E}\rightarrow TM$ defines a Lie algebra homomorphism on
sections, i.e.,
\begin{equation}\label{lah}
\rho_\ast([s_1,s_2])=[\rho_\ast(s_1),\rho_\ast(s_2)]\,, \qquad
\forall s_1,s_2\in \Gamma(\mathcal{E})\, ,
\end{equation}
where the brackets in the r.h.s. stand for the commutator of
vector fields.

It is instructive to look at the local coordinate expression of
the above relations.  Let $x^\mu $ be a coordinate system on a
trivializing chart $\mathcal{U}\subset M$ and let $s_a$ be a frame
of $\mathcal{E}|_\mathcal{U}$. By definition, we have
\begin{equation}\label{Xf}
    [s_a,s_ b]=f_{ab}^c(x) s_c \,,
    \qquad \rho_\ast(s_a)=X_a^\mu(x)\partial_\mu\,,
\end{equation}
$$
\mu =1,..., \dim
M\,,\;\;\;\;\;\;a=1,...,\mathrm{rank}\,\mathcal{E}\,.
$$
In view of Rels. (\ref{lah}) and (\ref{anc}) the structure
functions $f_{ab}^c, X_a^\mu\in C^{\infty}(U)$ meet the following
conditions:
\begin{equation}\label{lah1}
[X_a,X_b]^\nu :=X_a^\mu\partial_\mu X_b^\nu - X_b^\mu\partial_\mu
X_a^\nu=f_{ab}^cX_c^\nu\,,
\end{equation}
\begin{equation}\label{anc1}
    f_{ab}^df_{dc}^e-X_c^\mu\partial_\mu
    f_{ab}^e+cycle(a,b,c)=0\,.
\end{equation}
Notice that the second relation is automatically satisfied for any
vector bundle  $\mathcal{E}$ of rank 1 or 2, whereas  in the case
of  ${\rm rank}\,\mathcal{E}>2$ it becomes an actual restriction
on the structure functions $f_{ab}^c$.

In general, $\rho(\mathcal{E})$ is not a smooth subbundle of $TM$
as the rank of the distribution $\rho (\mathcal{E})$ may vary from
point to point. Nonetheless, in view of (\ref{lah}),
$\rho(\mathcal{E})$ generates a (singular) integrable distribution
in the sense of Sussman \cite{Sussman}: for each $p\in M$ there is
a smooth submanifold $\Sigma_p\subset M$ such that $p\in\Sigma_p$
and  $T_q\Sigma_p = \rho(\mathcal{E}_q)$ for any $q\in \Sigma_p$.
 The corresponding
foliation will be denoted by ${F}(M)$.

\vspace{5mm} \noindent {\textit{Example.}} Any tangent bundle $TM$
may be viewed as a Lie algebroid with the Lie bracket given by the
commutator of vector fields and the anchor $\rho=\mathrm{id}: TM
\rightarrow TM$.

\subsection{Differential geometry of Lie algebroids.}

One can regards the concept of a Lie algebroid as a tool for
transferring all the usual differential-geometric constructions
from a tangent bundle to an abstract vector bundle. In particular,
it is possible to define the Lie-algebroid counterpart of the
exterior calculus. Denote by $\Lambda(\mathcal{E})=\oplus
\Lambda^p(\mathcal{E})$ the exterior algebra of sections
$\Gamma(\wedge^\bullet\mathcal{E}^\ast)$, $\mathcal{E}^\ast$ being
the bundle dual to $\mathcal{E}$. Consider the following nilpotent
operator $d:\Lambda^p(\mathcal{E})\rightarrow
\Lambda^{p+1}(\mathcal{E})$:
\begin{equation}\label{d}
d\alpha(s_0,...,s_p)=\sum_{k=0}^{p}(-1)^k\rho_\ast(s_k)
(\alpha(s_0,...,\hat{s}_k,...,s_p))
\end{equation}
$$
+\sum_{k<n=1}^p(-1)^{k+n}\alpha([s_k,s_n],s_0,...,
\hat{s}_k,...,\hat{s}_n,...,s_p)\, ,
$$
for all $s_0, s_1, \ldots , s_p \in \Gamma (\mathcal{E})$. Since
$d^2=0$, we have a generalization of the De Rham complex. We will
refer to elements of $\Lambda^p(\mathcal{E})$ as
$\mathcal{E}$-$p$-forms, or just $p$-forms when it cannot lead to
confusion. Note that $\Lambda^0(\mathcal{E})$ is naturally
identified with $C^\infty(M)$.

More generally, one may consider the tensor product
$\mathcal{E}\otimes V$, where $V\rightarrow M$ is a vector bundle
with connection $\nabla$. Then, $\nabla$  induces the covariant
derivative $\nabla_\rho: \Lambda^p(\mathcal{E},V)\rightarrow
\Lambda^{p+1}(\mathcal{E},V)$ on the space
$\Lambda(M,\mathcal{E})=\Lambda(\mathcal{E})\otimes \Gamma(V)$ of
$\Gamma(V)$-valued $\mathcal{E}$-forms:
\begin{equation}\label{}
    \nabla_\rho
    \omega(s_0,...,s_p)=\sum_{k=0}^p(-1)^k\nabla_{\rho_\ast(s_k)}(\omega(s_0,...,\hat{s}_k,...,s_p))
\end{equation}
$$
+\sum_{k<n=1}^p(-1)^{k+n}\alpha([s_k,s_n],s_0,...,\hat{s}_k,...,\hat{s}_n,...,s_p)\,.
$$
The curvature of $\nabla_\rho$ is defined in the usual way:
\begin{equation}\label{curv}
    R=\nabla^2_\rho: \Gamma(V)\rightarrow
    \Lambda^2(\mathcal{E},V)\,.
\end{equation}
One may verify that
\begin{equation}\label{}
    R(fu)=fRu\,,\;\;\;\;\;\forall f\in C^\infty(M)\,,\; \forall u\in
    \mathcal{E}\,,
\end{equation}
so that in each coordinate chart the curvature $R$ is given by a
matrix valued 2-form determining a $C^\infty(M)$-linear
automorphism of $\Gamma(V)$. Like the curvature of the bundle
connection $\nabla$, $R$ satisfies the \textit{Bianchi identity}
\begin{equation}\label{bi}
 [\nabla_\rho, \nabla^2_\rho]=0  \Leftrightarrow   \nabla_\rho R=0\,.
\end{equation}
(To write the last formula we extend the action of $\nabla$ from
$V$ to the tensor product $V\otimes V^\ast$ by the usual formulas
of differential geometry.)

In what follows we will mostly deal with the case $V=\mathcal{E}$.
Then, in addition to the curvature, one more covariant of the
connection can be introduced.  The \textit{torsion} $T$ of a Lie
algebroid connection $\nabla_\rho$ is an element of
$\Lambda^2(\mathcal{E},\mathcal{E})$ defined by the rule
\begin{equation}\label{tor}
   \Gamma(\mathcal{E}) \ni T(u,v)=\nabla_{\rho_*(u)}v-\nabla_{\rho_*(v)}u-
   [u,v]\,,\;\;\;\;\; \forall \,u,v
   \in \Gamma (\mathcal{E})\,.
\end{equation}

If $\Gamma_{\mu b}^a$ are coefficients of the connection $\nabla$
with respect to local coordinates $x^\mu$ and a frame $s_a$, then
the components of the torsion tensor read
\begin{equation}\label{tor1}
    T_{ab}^c =X_a^\mu\Gamma_{\mu b}^c-X_b^\mu\Gamma_{\mu
    a}^c-f_{ab}^c\,.
\end{equation}
The components of the curvature tensor $R$ are
\begin{equation}\label{rdef1}
R_{abc}^d=X_a^\mu X_b^\nu R_{\mu\nu
c}^d\,,
\end{equation}
where
\begin{equation}\label{rdef}
    R_{\mu\nu c}^d=\partial_\mu\Gamma_{\nu c}^d -\partial_\nu
    \Gamma_{\mu c}^d +\Gamma_{\mu a}^d\Gamma_{\nu c}^a-\Gamma_{\nu
    a}^d\Gamma_{\mu c}^a
\end{equation}
is the curvature of $\nabla$.  There is a simple formula relating
the exterior and covariant derivatives:
\begin{equation}\label{dn}
d\alpha(s_0,...,s_p)=\sum_{k=0}^{p}(-1)^k(\nabla_{\rho_\ast(s_k)}\alpha)(s_0,...,\hat{s}_k,...,s_p)
\end{equation}
$$
+\sum_{k<n=1}^p(-1)^{k+n}\alpha(T(s_k,s_n),s_0,...,\hat{s}_k,...,\hat{s}_n,...,s_p)\,;
$$
here we use the isomorphism $
\Lambda^p(\mathcal{E})\simeq\Lambda^0(\mathcal{E},\wedge^p
\mathcal{E})$. A straightforward computation yields the
\textit{torsion Bianchi identity }
\begin{equation}\label{tbi}
\nabla_{c}T^d_{ab}+R^d_{abc}+cycle(a,b,c)=0\,,
\end{equation}
where $\nabla_a:=\nabla_{\rho_*(s_a)}$.

In this paper we are interested in the Lie algebroids endowed with
a closed and non-degenerate 2-form $\omega\in
\Lambda^2(\mathcal{E})$. A 2-form $\omega$ is called
non-degenerate if the equality
\begin{equation}\label{}
    \omega(u,v)=0\,,\;\;\;\;\;  \forall u\in
    \Gamma(\mathcal{E})\,,
\end{equation}
implies $v=0$. In terms of local coordinates the closedness
condition $d\omega=0$ reads
\begin{equation}\label{clos}
    X_c^\mu\partial_\mu\omega_{ab}+\omega_{cd}f_{ab}^d+cycle(a,b,c)=0\,,
\end{equation}
where $\omega_{ab}:=\omega(s_a,s_b)$. Extending the analogy with
classical differential geometry, we refer to $\omega$ as the
\textit{symplectic form} and call the triple
$(\mathcal{E},\rho,\omega)$  the \textit{symplectic Lie
algebroid}\footnote{This is a particular example of \textit{
triangular Lie bialgebroids} studied in Ref. \cite{MX}.}.

\subsection{Quasi-symplectic manifolds.} It is well known that any
symplectic structure on a Lie algebroid $\mathcal{E}\rightarrow M$
gives rise to a Poisson structure on the base manifold $M$. It is
this Poisson structure we are going to quantize by the BFV-BRST
method.

\vspace{5mm}

\noindent{\bf Proposition 2.1.} \textit{Let  $(\mathcal{E},\rho,
\omega)$ be a symplectic Lie algebroid, then $M$ is a Poisson
manifold  w.r.t. to the following Poisson bracket:
\begin{equation}\label{br}
\{f,g\}=\omega^{-1}(df,dg)\,,\;\;\;\;\;\forall f,g\in
C^{\infty}(M)\simeq\Lambda^0(\mathcal{E})\,;
\end{equation}
here $\omega^{-1}$ is the bi-section inverse to the symplectic
form $\omega\in \Lambda^2(\mathcal{E})$ and $d f , \, d g \in
\Lambda^1 (\mathcal{E}) $ are the differentials defined by
(\ref{d}).}

\vspace{5mm} \noindent{\bf Proof.} In terms of local coordinates
the Poisson bi-vector determining the bracket (\ref{br}) has the
form
\begin{equation}\label{biv}
    \alpha = \omega^{ab}X_a\wedge X_b \, \in \wedge^2TM  , \qquad \{f,g\}=
    \alpha^{\mu\nu} \partial_\mu f \partial_\nu g \, ,
\end{equation}
where $ X_a:=X_a^\mu\partial_\mu$,
$\omega^{ac}\omega_{cb}=\delta^a_b$, and
$\alpha^{\mu\nu}=\omega^{ab}X_a^\mu X_b^\nu$. The Jacobi identity
for $\alpha$ follows immediately from the Lie algebroid relations
(\ref{lah1}) and the closedness condition (\ref{clos}). Indeed,
using the Leibniz rule for the Schouten bracket of $\alpha$ with
itself, we get
\begin{equation}\label{alji}
\begin{array}{c}
   \frac14 [\alpha, \alpha]=\omega^{ab}\omega^{cd} X_a\wedge[X_b,X_c]\wedge X_d+
    \omega^{ab}[X_b,\omega^{cd}]X_a\wedge X_c \wedge X_d
    \\[3mm]
 =(\omega^{am} f_{mn}^b\omega^{nc}+\omega^{am}X_m^\mu
    \partial_\mu\omega^{bc})X_a\wedge X_b\wedge X_c=(d\omega)_{abc} X^a\wedge X^b\wedge X^c=0\,.
\end{array}
\end{equation}
The indices are lowered and raised with the help of the symplectic
form $\omega$ and its inverse.

\vspace{5mm} \noindent Since the rank of the anchor distribution
may vary through $M$, the induced Poisson structure (\ref{biv}) is
irregular in general (though it involves a non-degenerate
bi-vector $\omega^{-1}  \in \wedge^2 \mathcal{E} $). For this
reason and following the terminology of the work \cite{Vaisman} we
refer to $(M,\alpha)$ as the \textit{quasi-symplectic manifold}.
Accordingly,  Rel.(\ref{br}) is said to define a
\textit{quasi-symplectic representation} for the Poisson bi-vector
$\alpha$.

Given a symplectic Lie algebroid, we have two (singular)
foliations on $M$: the anchor foliation ${F}(M)$ and the
symplectic foliation ${S}(M)$ associated with the induced Poisson
structure  (\ref{biv}). Clearly, the latter foliation is
subordinated to the former one in the sense that any symplectic
leaf belongs to a leaf of the anchor foliation.

\vspace{5mm} \noindent {\it Remark.} A natural question to ask is
as follows: Given a Poisson bi-vector (\ref{biv}), where
$\omega_{ab}$ is some non-degenerate 2-form and $X_a$ is an
integrable distribution, are these data sufficient to define a
symplectic Lie algebroid? In general, the answer is negative,
since we do not require the local vector fields $X_a$ to be
linearly independent. Nonetheless, if $X$'s are linearly
independent on an everywhere dense domain in $M$, the answer is
positive. In that case the structure equations (\ref{anc1}) and
(\ref{clos})   follow immediately from the Jacobi identities for
the Schouten commutators of the local vector fields (\ref{lah1})
and the Poisson bi-vector (\ref{biv}). We will discuss this
question in more detail in Sect. 5 .

\subsection{Symplectic connection and curvature.}
The deformation quantization of quasi-symplectic Poisson manifolds
to be developed in the next sections involves one more geometric
ingredient, a \textit{symplectic connection}. This is defined as a
torsion-free Lie-algebroid connection respecting a symplectic
2-form, i.e.,
\begin{equation}\label{comc}
    \nabla_\rho \omega=0\,.
\end{equation}
Here we consider $\omega$ as a section of
$\Lambda^0(\mathcal{E},\wedge^2 \mathcal{E})$.

\vspace{5mm}

\noindent {\bf Proposition 2.2.} \textit{Any symplectic Lie
algebroid admits a symplectic connection.}

\vspace{5mm}

\noindent {\bf Proof.} We are looking for a symplectic connection
of the form $\nabla+\Delta\Gamma$, where $\nabla$ is an arbitrary
connection and $\Delta\Gamma\in
\Gamma(\mathcal{E}\otimes\mathcal{E}^\ast\otimes\mathcal{E}^\ast)$.
In terms of local coordinates the compatibility condition
(\ref{comc}) reads
\begin{equation}\label{sc}
    \nabla_c \omega_{ab}=\Delta\Gamma_{cab} - \Delta\Gamma_{cba}\,,
\end{equation}
where $\Delta \Gamma_{abc} =\Delta\Gamma^d_{ab}\omega_{dc}$.
Obviously, these equations cannot have a unique solution: any
tensor $\Delta\Gamma'_{abc}$, symmetric in $bc$, satisfies the
homogeneous equation and therefore it can be added to a given
solution $\Delta\Gamma_{abc}$ to produce another one. A particular
solution to Eq.(\ref{sc}) is given by
\begin{equation}\label{}
\Delta\Gamma_{ab}^c=-\frac12 \omega^{cd}\nabla_a \omega_{db}\,,
\end{equation}

Now let $\nabla_\rho$ be an arbitrary Lie-algebroid  connection
which respects $\omega$ and has torsion $T$. By making use of the
aforementioned ambiguity, one can define the new connection
$\nabla'_\rho=\nabla+\Delta\Gamma'$,
\begin{equation}\label{g1}
\Delta\Gamma'_{abc}=-\frac13(T_{abc}+T_{acb})\, , \qquad
T_{abc}=T_{ab}^d\omega_{dc}\,,
\end{equation}
which is also compatible with $\omega$. By definition
(\ref{tor1}), we have
\begin{equation}\label{t1}
    T'{}_{ab}^c=T_{ab}^c+\Delta\Gamma'{}_{ab}^c-\Delta\Gamma'{}_{ba}^c\,.
\end{equation}
Substituting (\ref{g1}) into (\ref{t1}) and lowering the upper
index with the help of $\omega$ we get
\begin{equation}\label{}
T_{abc}'=T_{acb}+cycle(a,b,c)=0\,.
\end{equation}
The last equality follows immediately from (\ref{dn}) with
$\omega$ in place of $\alpha$. Thus,  $\nabla'_\rho$ is a
symplectic connection.

\vspace{5mm} Let $R_{abc}^d$ be the curvature of a symplectic
connection. By analogy with Riemannian geometry we can define the
\textit{covariant curvature tensor} just lowering upper index with
the help of the  symplectic 2-form:
$R_{abcd}=R_{abc}^n\omega_{nd}$. The following symmetry properties
take place:
    \begin{equation}\label{sym}
    R_{abcd}=-R_{bacd}\,,\qquad
    R_{abcd}=R_{abdc}\,,\qquad
    R_{abcd}+R_{bcad}+R_{cabd}=0\,.
    \end{equation}
The first equality is obvious, the second one follows from the
definition (\ref{curv}) and the fact that $\nabla_\rho$ respects
$\omega$, the third equality is just the Bianchi identity
(\ref{tbi}).

\subsection{Examples.} Let us give some examples of symplectic Lie
algebroids and the corresponding quasi-symplectic Poisson
brackets. More examples of Lie algebroids, with or without
symplectic structure, can be found in \cite{Vaisman}, \cite{WCdS}.

\vspace{5mm}

\noindent{\it Example 1}. Any symplectic manifold $(M,\omega)$
gives rise to the symplectic Lie algebroid
$(TM,{\rm{id}},\omega)$. The quasi-symplectic Poisson structure is
given by
\begin{equation}\label{cqsr}
    \alpha=(\omega^{-1})^{\mu\nu}\partial_\mu\wedge\partial_\nu\,.
\end{equation}

To get a less trivial quasi-symplectic representation for $\alpha$
consider an almost complex structure $J$ compatible with $\omega$.
Recall that an almost complex structure is a smooth field of
automorphisms $J:TM\rightarrow TM$ obeying conditions
\begin{equation}\label{}
    J^2=-{\rm{id}}\;,\qquad \omega(JX,JY)=\omega(X,Y)\,,
\end{equation}
where $X,Y$ are arbitrary vector fields. $J$ being non-degenerate,
we get a quasi-symplectic representation for the Poisson bi-vector
$\alpha$ associated with the symplectic Lie algebroid
$(TM,J,\omega)$:
\begin{equation}\label{}
    \alpha=(\omega^{-1})^{\mu\nu}J_\mu\wedge J_\nu\,.
\end{equation}
Here $J=dx^\mu J_\mu^\nu\partial_\nu$ and
$J_\mu=J_\mu^\nu\partial_\nu$.

\vspace{5mm}

\noindent {\it Example 2.} Generalizing previous example, consider
a pair of Schouten-commuting bi-vectors $\beta$ and $\omega$,
where  the former is a Poisson one and the latter is
non-degenerate. The triple $(T^\ast M,\beta,\omega)$ defines a
symplectic Lie algebroid with the structure functions
$$
[\beta(dx^\mu),\beta(dx^\nu)]=\partial_\lambda\beta^{\mu\nu}\beta(dx^\lambda)\,,\;\;\;\;\;
\beta(dx^\mu)=\beta^{\mu\nu}\partial_\nu\,.
$$
The induced Poisson structure on $M$ is given by
\begin{equation}\label{}
\alpha=(\omega^{-1})_{\mu\nu}\beta^{\mu\gamma}\beta^{\nu\lambda}\partial_\gamma
\wedge\partial_\lambda\,.
\end{equation}

\vspace{5mm} \noindent{\it Example 3}. Recall that a Lie algebra
$L$ is called \textit{quasi-Frobenius} \cite{El}  if it admits a
non-degenerate central extension $L_c$ of the form
$$
[p_a,p_b]=f^d_{ab}p_d+\omega_{ab}c\,,\;\;\;\;\;
[c,p_a]=0\,,\;\;\;\det(\omega_{ab})\neq 0\,.
$$
The Jacobi identity for $L_c$ requires the  non-degenerate matrix
$\omega_{ab}$, determining the central extension,  to be a
2-cocycle of the Lie algebra $L\simeq L_c/c$.

Given an action  $\rho: L\rightarrow {\rm Vect}(M)$ of the Lie
algebra $L$ on $M$ by smooth vector fields $X_a=\rho(p_a)$, one
can define a symplectic Lie algebroid associated with the trivial
vector bundle $M \oplus L$, anchor $\rho$, and symplectic form
$\omega_{ab}$. The induced quasi-symplectic structure on $M$ reads
\begin{equation}\label{}
  \alpha=\omega^{ab}X_a\wedge
  X_b\,,\;\;\;\;\;\;\;\;\;\omega^{ac}\omega_{cb}=\delta^a_b\,.
\end{equation}
A simple quantization procedure for such Poisson brackets has been
proposed in \cite{DILSh}.

\vspace{5mm}

\noindent{\it Example 4.} Let $(M,\alpha)$ be a 2-dimensional
Poisson manifold. We say that the bi-vector $\alpha$ is
\textit{quasi-homogeneous} if there exist a volume form $\omega$
and a vector field $Y$ such that the function $h=\omega(\alpha)$
obeys condition $Yh=h$.

It turns out that any quasi-homogeneous Poisson manifold is also a
quasi-symplectic one. Namely, a simple computation yields
\begin{equation}\label{qhc}
\alpha=X\wedge Y\,,\;\;\;\;\;\;\; [X,Y]=(1-{\rm div}_\omega Y)X\,.
\end{equation}
Here $X$ is the Hamiltonian vector field associated with the
Hamiltonian $h$ and the symplectic (volume) form $\omega$. The
structure equations (\ref{anc1}) and (\ref{clos}) are
automatically satisfied by the reason of dimension and we get a
symplectic Lie algebroid associated with the trivial vector bundle
$M\oplus \mathbb{R}^2$.

For instance, the following polynomial Poisson brackets on 2-plane
\begin{equation}
\{x,y\}=x^my^n+x^ky^l\,,
\end{equation}
$$
\delta=nk-lm\neq 0\,,\;\;\;\;m,n,k,l\in \mathbb{N}\,,
$$
are quasi-homogeneous w.r.t. $\omega=dx\wedge dy$ and
\begin{equation}
\begin{array}{l}
Y= \delta^{-1}(n-l)x\partial_x + \delta^{-1}(k-m)y
\partial_y\,,\\[3mm]
X=(mx^{m-1}y^n+kx^{k-1}y^l)\partial_y
-(nx^my^{n-1}+lx^ky^{l-1})\partial_x\,.
\end{array}
\end{equation}
In accordance  with (\ref{qhc})
\begin{equation}\label{ab}
[Y,X]=(1-\delta^{-1}(n+k-l-m))X\,,
\end{equation}
and we arrive at the symplectic Lie algebroid associated with the
two-dimensional quasi-Frobenius Lie algebra (\ref{ab}) (see the
previous example).

\subsection{Counter-example.}

As we have seen the quasi-symplectic manifolds constitute a wide
class of Poisson manifolds. Here is an example  of a Poisson
manifold that does not admit any quasi-symplectic representation
(even locally).

\vspace{5mm} \noindent {\bf Proposition 2.3.} {\it The Lie-Poisson
bracket on the dual of $so(3)$ algebra does not admit a
quasi-symplectic representation.}

\vspace{5mm}\noindent {\bf Proof.} The Poisson bracket in
question is of the form
\begin{equation}\label{}
    \{x^i,x^j\}=\sum_{k}\epsilon^{ijk}x^k\,,
\end{equation}
where $x^i$ are linear coordinates in $\mathbb{R}^3$ and
$\epsilon^{ijk}$ is the Levi-Civita tensor. The only irregular
point is $0\in \mathbb{R}^3$, where the  rank of the Poisson
bracket is equal to zero; at the other points the rank equals 2.
The leaves of the symplectic foliation ${S}(\mathbb{R}^3)$ are
exactly the level sets of the Casimir function
$(x^1)^2+(x^2)^2+(x^3)^2$, i.e., spheres centered at the origin.

Since any vector bundle $\mathcal{E}$ over $\mathbb{R}^3$ is
trivial, we may look for an anchor being just an integrable vector
distribution $X_a\in {\rm Vect}(\mathbb{R}^3)$. For the same
reason, any symplectic 2-$\mathcal{E}$-form is given by an
invertible skew-symmetric matrix $\omega_{ab}(x)$ on
${\mathbb{R}}^3$. Clearly, each sphere from ${S}(\mathbb{R}^3)$ is
entirely contained in some leaf of the anchor foliation
${F}(\mathbb{R}^3)$; so, we write: ${S}(\mathbb{R}^3)\subset
{F}(\mathbb{R}^3)$. The existence of a quasi-symplectic
representation is expressed by the equality
\begin{equation}\label{ce}
X^i_a\omega^{ab}X_b^j=\sum_k\epsilon^{ijk}x^k\,.
\end{equation}
The key to the analysis of this equation lies with the rank $r$ of
the vector distribution at the origin. \textit{A priory}, $r$ may
take any value from 0 to 3. Let us show that any assumption about
$r$ leads to a contradiction.

\vspace{3mm}\noindent {\it The case}$\; r=0$: This possibility is
ruled out by comparing the order of zero on both sides of the
equality (\ref{ce}). Indeed, since all $X$'s must vanish at $x=0$,
the order of zero on the l.h.s. is of at least 2, while the r.h.s.
tends to zero linearly.

\vspace{3mm}\noindent {\it The cases} $\;r=1,2$: There is an
integral leaf of dimension 1 or 2 passing through the origin and
intersecting \textit{transversally} at least one of the symplectic
spheres. (Otherwise, this leaf would  be entirely contained in one
of the spheres, and thus, could not reach the origin.) But this
contradicts to the inclusion ${S}(\mathbb{R}^3)\subset
{F}(\mathbb{R}^3)$.

\vspace{3mm}\noindent {\it The case} $\;r=3$: Passing, if
necessary, to another basis we may assume that
$X_a=(X_i,X_\alpha)$, where $X_i=\partial_i$ and $X_\alpha=0$.
Then the matrix $\omega^{ab}$ takes the block form
\begin{equation}\label{block}
\left(
\begin{array}{cc}
  \omega^{ij} &  \omega^{i\beta}\\
  \omega^{\alpha j} & \omega^{\alpha\beta} \\
\end{array}
\right)
\end{equation}
with $\omega^{ij}=\sum \epsilon^{ijk}x^k$. Among various equations
on the matrix elements of (\ref{block}), expressing the fact of
closedness of $\omega$, one can find the following one:
\begin{equation}\label{}
    \omega^{\alpha k}\partial_k\omega^{ij}
    +\omega^{ik}\partial_k\omega^{j\alpha}-\omega^{jk}\partial_k\omega^{i\alpha}=0\,.
\end{equation}
Since $\omega^{ij}(0)=0$ and
$\partial_k\omega^{ij}(0)=\epsilon^{ijk}$, the last equation
implies that $\omega^{\alpha i}(0)=0$, and hence the entire matrix
(\ref{block}) must degenerate at the origin. This contradiction
concludes the proof.

\section{Poisson description of symplectic Lie algebroids}

In order to construct as well as physically interpret the
deformation quantization of quasi-symplectic manifolds it is
convenient to think of $(M,\alpha)$ as the phase space of some
(gauge invariant) mechanical system with zero Hamiltonian. In what
follows we will use the standard terminology from  the theory of
constrained systems: first and second class constraints,
gauge-fixing conditions, ghost variables,  BRST charge etc.
\cite{HT}. It should be noted  that unlike the common practice we
will consider Hamiltonian constraints that are defined by a
\textit{section} of some vector bundle $\mathcal{E}\rightarrow M$
rather than scalar functions on $M$. To provide the covariance of
the quantization with respect to the bundle automorphisms an
appropriate linear connection is needed, and that requires some
modification of the conventional BRST formalism
\cite{BFV,BFV2,HT}. In particular, it will be convenient to use
non-canonical commutation relations for ghost variables. The
details will be explained below.

Now let us outline the basic steps of our approach. The main idea
is to quantize a quasi-symplectic manifold $M$ by means of its
suitable embedding into a certain supermanifold endowed with ``a
more simple'' Poisson structure.  The construction of such an
embedding involves a quite standard machinery of the Hamiltonian
BRST theory \cite{BFV2,HT,BT}; it can be subdivided into three
steps. First, using the Lie algebroid structure, we represent
$(M,\alpha)$ as a second-class constrained system on the vector
bundle $\mathcal{E}^\ast$ dual to the Lie algebroid $\mathcal{E}$.
As the next step, the second-class constrained system on
$\mathcal{E}^\ast$ is converted into an equivalent gauge system on
the direct sum of vector bundles $\mathcal{N}=\mathcal{E}^*\oplus
\mathcal{E}$. The equivalence just means that the Poisson algebra
of physical observables on $\mathcal{N}$ is isomorphic to the
Poisson algebra of smooth functions on $(M, \alpha)$. Finally, the
classical gauge system is covariantly quantized by the BFV-BRST
method. The key point is that the space of physical observables on
$\mathcal{N}$, being identified with a certain BFV-BRST cohomology
in ghost number zero, carries a simple Poisson structure which can
easily be quantized. By construction, the associative product on
the algebra of quantum observables on $\mathcal{N}$ induces a
$\ast$-product on the original quasi-symplectic manifold
$(M,\alpha)$.

For the case of symplectic manifolds, including second-class
constrained system, such a program was first implemented in
\cite{GL}, \cite{BGL} establishing detailed correspondence between
the key ingredients of the BRST theory and the Fedosov deformation
quantization.

\subsection{Symplectic embedding.}
We start with the  description of a symplectic embedding of
$(M,\alpha)$ into the dual bundle of the corresponding Lie
algebroid. It is well known that $\mathcal{E}^*$ carries a natural
Poisson structure, which is dual to the Lie algebroid structure
\cite{WCdS}, \cite{Courant}. A proper modification of this Poisson
structure in the presence of a symplectic 2-form is offered by the
next proposition.

\vspace{5mm} \noindent {\bf Proposition 3.1.} {\it Let
$(\mathcal{E}, \rho, \omega)$ be a symplectic Lie algebroid
corresponding to a quasi-symplectic manifold $(M,\alpha)$. Then
$C^\infty(\mathcal{E}^\ast)$ can be equipped  with the following
Poisson brackets:
\begin{equation}\label{pbr}
\{x^\mu,x^\nu\}=0\,,\qquad \{p_a,x^\mu\}=X_{a}^\mu\,,\qquad
\{p_a,p_b\}=f_{ab}^cp_c+\omega_{ab}\,.
\end{equation}
Here $x^\mu$ are local coordinates on $M$  and $p_a$ are linear
coordinates on the fibers of $\mathcal{E}^\ast$. The Poisson
manifold $(M,\alpha)$ is symplectically imbedded into
$\mathcal{E}^\ast$ as zero section.}

\vspace{5mm}

\noindent{\it Remark.} Although the definition of the brackets on
$\mathcal{E}^\ast$ involves local coordinates, the Poisson
structure (\ref{pbr}) is actually coordinate independent, so the
relationship  between the Lie algebroid structure on $\mathcal{E}$
and the Poisson bi-vector on $\mathcal{E}^\ast$ is intrinsic. The
Jacobi identity for (\ref{pbr}) generates the full set of the Lie
algebroid axioms as well as the closedness condition for the
symplectic structure.

\vspace{5mm} In terms of local coordinates $(x^\mu, p_a)$ one may
identify the base manifold $M$ with those points of
$\mathcal{E}^\ast$ for which
\begin{equation}\label{pa}
    p_a=0\,.
\end{equation}
Since
\begin{equation}\label{}
  \det (\{p_a,p_b\} )|_{p=0}=\det(\omega_{ab})\neq 0\,,
\end{equation}
the canonical imbedding $M\hookrightarrow\mathcal{E}^\ast$ defined
by (\ref{pa}) is symplectic, and the induced Poisson structure on
$M$ reads
\begin{equation}\label{}
\{f,g\}=\omega^{ab}(X_a^\mu\partial_\mu f)(X_b^\nu\partial_\nu
g)\,,\;\;\;
  \;\;\;\;\forall\,f,g
  \in C^{\infty}(M)\,.
\end{equation}
\ From the physical viewpoint, this bracket can be thought of as
the Dirac bracket associated with the second-class constraints
(\ref{pa}), where $f$ and $g$ are taken to be $p$-independent
functions on $\mathcal{E}^\ast$.

\subsection{Classical conversion.}
Choosing a symplectic connection $\nabla_\rho$, one can extend the
Poisson structure (\ref{pbr}) on $\mathcal{E}^*$ to that on the
direct sum $\mathcal{N}=\mathcal{E}\oplus \mathcal{E}^\ast$.
Namely, if $y^a$ are linear coordinates on the fibers of
$\mathcal{E}$, then the corresponding Poisson brackets read
\begin{equation}\label{br3}
    \begin{array}{ll}
      \{x^\mu,x^\nu\}=0\,,& \{p_a,x^\mu\}=X^\mu_a(x)\,, \\[2mm]
      \{x^\mu,y^a\}=0\,,& \{p_a,y^b\}=-\Gamma_{ac}^b(x)y^c\,, \\[2mm]
      \{y^a,y^b\}=\omega^{ab}(x)\,,\qquad&\{p_a,p_b\}=\omega_{ab}(x)+
      f_{ab}^c(x)p_c-\frac12R_{abcd}(x)y^cy^d\,.
    \end{array}
\end{equation}
Here $\Gamma^c_{ab}$  are the coefficients of the connection
$\nabla_{\rho}$ and $R_{abcd}$ is the corresponding curvature
tensor.

The brackets (\ref{br3}) are well-defined and meet the Jacobi
identity. Verifying the Jacobi identity, one gets the
compatibility condition (\ref{comc}), the definition of the
curvature tensor (\ref{rdef1}), the Bianchi identity (\ref{bi}),
and the axioms of a symplectic Lie algebroid.

Now we aim to replace the second-class constrained system
(\ref{pbr}), (\ref{pa}) on $\mathcal{E}^*$ with an equivalent
gauge system on the extended Poisson manifold $\mathcal{N}$. In
the Hamiltonian formalism a reparametrization invariant gauge
system is completely specified by a set of first class constraints
$T_a=0$ defining some coisotropic submanifold $\Sigma\subset
\mathcal{N}$ (a constraint surface). The quotient of $\Sigma$ by
the Hamiltonian action of $T$'s is assumed to be isomorphic to the
quasi-symplectic manifold $(M,\alpha)$ and this is the sense in
which the equivalence will be established between the original
Poisson manifold and the effective gauge theory. In fact, for the
purposes of deformation quantization it is sufficient to work with
a \textit{formal gauge system} on $\mathcal{N}$ in the sense that
the first class constraints $T_a$ are allowed to be given by
formal power series in $y$'s. It is required, however, that the
canonical projection of the formal coisotropic submanifold
$\Sigma$ onto $\mathcal{E}^\ast$ to coincide with the well-defined
constraint surface (\ref{pa}), i.e., with $M$. This allows one to
assign a precise meaning to the Hamiltonian reduction by the
formal first class constraints.

Thus, we are looking for a set of Hamiltonian constraints
$T_a(x,p,y)$ obeying conditions
\begin{equation}\label{conv}
\{T_a,T_b\}=U_{ab}^cT_c\,,\qquad
T_a(x,p,y)|_{\mathcal{E}}=T_a(x,p,0)=p_a\,,
\end{equation}
where $U_{ab}^c(x,p,y)$ are some structure functions.
Geometrically, one can thought of $T$'s as a section of the vector
bundle $\pi : \mathcal{E}\oplus \mathcal{E}^*\rightarrow
\mathcal{E}$ over the base $\mathcal{E}$, with $\pi$ being the
canonical projection onto the first factor.

\vspace{5mm} \noindent \textbf{Proposition 3.2.}\textit{The
equations (\ref{conv}) have a solution of the form}
\begin{equation}\label{ans}
T_a=p_a+\sum_{n=1}^\infty T_a^n\,,\;\;\;\;\;\;T_a^n=t_{ab_1\cdots
b_n}(x)y^{b_1}\cdots
    y^{b_n}\,,
\end{equation}
\textit{where the coefficients $t_{ab_1\cdots b_n}(x)$ do not
depend on $p$'s.}

\vspace{5mm}
 \noindent {\it Remark.} In the physical literature,
the passage from a given second-class constrained system to an
equivalent first-class one is known as the \textit{conversion
procedure}; accordingly, $y$'s are called \textit{conversion
variables}. In the local setting, i.e., for a sufficiently small
domain in the extended phase space, the existence of the
conversion is ensured by a fairly general theorem \cite{BT}.
Moreover, passing, if necessary, to an equivalent basis of second
class constraints, it is possible to have a solution with
$U_{ab}^c=0$ (abelian conversion). Here, however, we concern with
account of global geometry that requires to consider a non-abelian
conversion in general.

\vspace{5mm} \noindent \textbf{Proof.} Substituting the expansion
(\ref{ans}) into the involution relations (\ref{conv}) and
extracting contribution to zero order in $y$'s, we find
\begin{equation}\label{}
(f_{ab}^c-U_{ab}^c)p_c+t_{an}\omega^{nm}t_{bm}+\omega_{ab}=0\,.
\end{equation}
A particular solution to this equation is obvious:
\begin{equation}\label{ps}
U_{ab}^c=f_{ab}^c\,,\qquad t_{ab}=-\omega_{ab}\,.
\end{equation}
Taking this solution, one gets the following chain of equations
for higher orders in $y$'s:
\begin{equation}\label{hord}
F_a^s:=\partial_{[a} T^{s+1}_{b]} - B^s_{ab} = 0\,,\;\;\;\;s\geq
1\,,
\end{equation}
where
\begin{equation}\label{}
\begin{array}{l}
B^1_{ab} = (X^i_{[a}\partial_i \omega_{b]c} +
\omega_{d[a}\Gamma_{b]c}^d
+ f_{ab}^d\omega_{dc})y^c  \,,\\[4mm]
B^2_{ab} =
\{p_{[a},T^2_{b]}\} + f_{ab}^cT^2_c -\frac12R_{abcd}y^cy^d\,,\\[4mm]
\displaystyle B^s_{ab} =
\{p_{[a},T^s_{b]}\}+f_{ab}^cT^s_c+\sum_{n=2}^s
\{T_a^n,T_b^{s+2-n}\}\,,\;\;\;\; s\geq 3\,.
\end{array}
\end{equation}
Hereinafter the square brackets denote anti-symmetrization of
indices and $\partial_a$ is the partial derivative with respect to
$y^a$. The form of the equations (\ref{hord}) suggests to
interpret $T^s_a$ as the components of 1-form $T^s=T^s_ady^a$
defined on the linear space of $y$'s. Thus, we can write
\begin{equation}\label{hord1}
F^s=dT^{s+1} - B^s=0\,,\;\;\;\;s\geq 1\,,
\end{equation}
where $d$ is the usual exterior differential with respect to
$y$'s, and $B^s$ is a given 2-form provided the 1-forms
$T^1,...,T^{s}$ have been already determined. According to the
Poincar\'e lemma,  Eqs.(\ref{hord1}) are consistent iff the
2-forms $B^s$ are closed. In this case, the general solution to
(\ref{hord}) reads
\begin{equation}\label{rf}
T_a^{s+1}=\frac{1}{s+2}y^bB_{ba}^s+\partial_a C^s\,,
\end{equation}
$C^s$ being an arbitrary monomial of degree $s+2$. The closedness
of  $B$'s is now proved by induction in $s$. Consider the identity
\begin{equation}\label{}
    \{\bar{T}^s_a,\bar{T}^s_b\}-f_{ab}^d\bar{T}_d^s=\sum_{n=0}^{s-1}F^n_{ab}+B^s_{ab}+
    \cdots\,,
\end{equation}
where $ \bar{T}_a^s=p_a+\sum_{n=1}^s T^n_a$ and dots stand for the
terms of order higher than $s$. Taking the Poisson bracket of this
relation with $\bar{T}^s_c$ and using the Jacobi identity
\begin{equation}\label{}
\{\{\bar{T}^s_{a},\bar{T}^s_{b}\},\bar{T}_{c}^s\}+cycle(a,b,c)=0\,,
\end{equation}
we can write
\begin{equation}\label{}
\begin{array}{l}
\left(f_{ab}^d \{\bar{T}^s_c,
\bar{T}_d^s\}+\{\bar{T}_c,f_{ab}^d\}\bar{T}_d\right)
dy^a\wedge dy^b\wedge dy^c = \\[4mm]
\displaystyle
=\left(\sum_{n=0}^{s-1}\{F^n_{ab},\bar{T}_c^s\}-\{B^s_{ab},\bar{T}_c^s\}+
    \cdots \right)dy^a\wedge dy^b\wedge dy^c\,.
\end{array}
\end{equation}
With account of (\ref{br3}) and the Lie algebroid relations
(\ref{anc1}), the contribution to the ($s-1$)-th order of the last
equation is given by
\begin{equation}\label{}
    dB^s=\left(f_{ab}^dF_{dc}^{s-1}+\sum_{n=0}^{s-1}\{F_{ab}^{n},T^{s-n+2}_c\}\right)dy^a\wedge
    dy^b\wedge dy^c\,.
\end{equation}
But the r.h.s. of this relation vanishes by the induction
hypothesis. Thus, $B^s$ is a closed 2-form and the recurrent
formula (\ref{rf}) gives the general solution for $T_a$.

Notice that the ambiguity concerning the choice of arbitrary
functions $C^s$, entering the general solution for $T_a$, can be
removed by imposing the $y$-transversality condition
\begin{equation}\label{yT}
    y^aT_a^s=0\,,\;\;\;\;s\geq 1\,.
\end{equation}
Then it follows from Eq. (\ref{rf}) that
$$
y^a\partial_a C^s (x,y)=(s+2)C^s(x,y)=0 \;\;\Rightarrow
\;\;C^s(x,y)=0\,.
$$

\vspace{5mm} \noindent {\it Remark.} For the case of symplectic
manifolds, the convergence of the series (\ref{ans}) in a tubular
neighborhood of $M$ was proved in \cite{EW} under assumptions of
analyticity and compactness. It seems that the same arguments are
applicable to any quasi-symplectic manifold provided all the
structure functions are real-analytical and $M$ is compact.

\vspace{5mm}

Now to see the equivalence of the constructed gauge system on
$\mathcal{N}$ to the original Hamiltonian system on $M$ it
suffices to note that equations $\chi^a:= y^a=0$ are well-defined
gauge-fixing conditions for the first class constraints $T_a=0$.
Indeed,
\begin{equation}\label{}
\det\left.\left(
\begin{array}{cc}
  \{T_a,T_c\}  & \{T_a,\chi^d\} \\
  \{\chi^b,T_c\}&  \{\chi^b,\chi^d\}
\end{array}
\right)\right|_{T=\chi=0}= \det\left(
\begin{array}{ccc}
   0  & -\delta_a^d \\
  \delta^b_c  & \omega^{bd}
\end{array}
\right)= 1\,.
\end{equation}
Therefore, the reduced Poisson manifold (physical phase space) is
isomorphic to the constraint surface  $T_a=\chi^b=0$. The last
equations are obviously  equivalent to $p_a=y^a=0$, i.e., defines
the canonical projection $\varphi: \mathcal{N}\rightarrow M$. The
explicit description of the resulting Poisson structure on $M$ can
be obtained by introducing the Dirac bracket for the second-class
constraints $(T_a,\chi^b)$. Identifying the space of smooth
functions on $M$ with the subspace of $p$- and $y$-independent
functions on $\mathcal{N}$, it is easy to see that $\varphi:
\mathcal{N}\rightarrow M$ is a Poisson map relating the Dirac
bracket $\{\cdot,\cdot\}_D$ on $\mathcal{N}$ with the initial
quasi-symplectic bracket on $M$, i.e.,
\begin{equation}\label{}
    \varphi_*(\{f,g\}_D)=\{\varphi_*(f),\varphi_*(g)\}_M\,.
\end{equation}

\section{Quantization}

Having realized the quasi-symplectic manifold $(M, \alpha)$ as a
formal gauge system on $\mathcal{N}$ we are ready to perform its
BRST quantization. As usual, this implies further enlargement of
the phase space of the system by ghost variables, constructing a
nilpotent BRST charge, and identifying physical observables with
certain BRST cohomology classes.

\subsection{Ghost variables and the classical BRST charge.}
With each first class constraint $T_a$ we associate the pair of
anticommuting (Grassman odd) ghost variables
$(\mathcal{C}^a,\mathcal{P}_b)$ subject to the canonical Poisson
bracket relations
\begin{equation}\label{ghbr}
    \{\mathcal{C}^a,\mathcal{P}_b\}=\delta^a_b\,,\;\;\;\;
    \{\mathcal{C}^a,\mathcal{C}^b\}=\{\mathcal{P}_a,\mathcal{P}_b\}=0\,.
\end{equation}
It is quite natural to treat $\mathcal{C}^a$ and $\mathcal{P}_b$
as linear coordinates on the fibers of the vector bundles $\Pi
\mathcal{E}$ and $\Pi\mathcal{E}^*$, respectively. Here by $\Pi$
we denote  the parity reversion operation: being applied to a
vector bundle it transforms the bundle into the super-vector
bundle with the same base manifold and transition functions, and
the fibers being the Grassman odd vector spaces. Thus, the phase
space of our gauge system is extended to the direct sum of
(super-)vector bundles $\mathcal{M}=\mathcal{N}\oplus
\Pi\mathcal{N}$. This geometric interpretation places the ghosts
on equal footing with the conversion variables $y$'s and suggests
the following extension of the Poisson structure from
$\mathcal{N}$ to $\mathcal{M}$:
\begin{equation}\label{ghcr}
    \{p_a, \mathcal{C}^b\}= -\Gamma_{ac}^b(x)
    \mathcal{C}^c\,,\;\;\;\;\;\;\;\;
    \{p_a,\mathcal{P}_b\}=\Gamma_{ab}^c (x) \mathcal{P}_c\,.
\end{equation}
The brackets of the ghosts with $x^\mu$ and $y^a$ are equal to
zero. To meet the Jacobi identity one has to modify the Poisson
brackets of $p$'s by ghost terms as follows
\begin{equation}\label{modif}
    \{p_a,p_b\}=\omega_{ab}(x)+
    f_{ab}^c(x)p_c-\frac12R_{abcd}(x)y^cy^d-R_{abc}^d(x)\mathcal{C}^c\mathcal{P}_d
    \,.
\end{equation}
The other Poisson brackets (\ref{br3}) remain intact.

\vspace{5mm}\noindent {\it Remark.} At this point we slightly
deviate from the usual line of the BRST scheme, where the ghost
variables are assumed to Poisson-commute with functions on the
original phase space ($\mathcal{N}$ in our case) and, in
particular, with the first class constraints. In principle, it is
possible to work with the canonical Poisson brackets for ghosts,
setting the r.h.s of (\ref{ghcr}) to zero and omitting the last
term in (\ref{modif}), but this leads to nonlinear transformations
of $p_a$ under bundle automorphisms ($p$'s are mixed with the
ghost bilinears $\mathcal{C}^a\mathcal{P}_b$). We refer to
\cite{GL} for the details of this construction in the case where
$M$ is a symplectic manifold (\ref{cqsr}). As we will see bellow,
these non-canonical Poisson brackets of ghosts can be naturally
incorporated into the BRST quantization procedure making it
explicitly covariant.

\vspace{5mm}

Let  $\mathcal{F}(M)$ denote the super-Poisson algebra of
functions on the  supermanifold $\mathcal{M}$; the elements of
$\mathcal{F}(\mathcal{M})$ are superfunctions of the
form\footnote{In what follows we omit the prefix ``super" whenever
possible.}
\begin{equation}\label{}
    A(x,p,y,\mathcal{C},\mathcal{P})=\sum A_{a_1\cdots
    a_k d_1\cdots d_n}^{b_1\cdots b_m}(x)y^{a_1}\cdots y^{a_k}
    \mathcal{C}^{d_1}\cdots\mathcal{C}^{d_n}
    \mathcal{P}_{b_1}\cdots\mathcal{P}_{b_m}\,,
\end{equation}
where $A_{a_1\cdots a_k d_1\cdots d_n}^{b_1\cdots b_m}(x)$ are
$\mathcal{E}$-tensors. In addition to the usual
$\mathbb{Z}_2$-grading, associated with the Grassman parity,
\begin{equation}\label{}
\epsilon(\mathcal{C}^a)=\epsilon(\mathcal{P}_b)=1\,,\;\;\;\;
\epsilon(x^i)=\epsilon(p_a)=\epsilon(y^b)=0 \;\;\;\;\; ({\rm
mod}\; 2)\,,
\end{equation}
the space $\mathcal{F}(\mathcal{M})$ is endowed with an additional
${\mathbb{Z}}$-grading by prescribing the following \textit{ghost
numbers} to the local coordinates:
\begin{equation}\label{}
{\rm gh}(\mathcal{C}^a)=1\,,\;\;\;\;{\rm
gh}(\mathcal{P}_a)=-1\,,\;\;\;\; {\rm gh}(x^i)={\rm gh}(p_a)={\rm
gh}(y^a)=0\,.
\end{equation}
The ghost number just counts the difference between powers of
$\mathcal{C}$'s and $\mathcal{P}$'s, entering homogeneous elements
of $\mathcal{F}(\mathcal{M})$, and is additive with respect to the
Poisson algebra operations:
\begin{equation}\label{}
    {\rm gh}(AB)={\rm gh}(A)+{\rm gh}(B)\,,\qquad {\rm gh}(\{A,B\})={\rm gh}(A)+
    {\rm gh}(B)\,.
\end{equation}
In particular, functions with zero ghost number form a subalgebra
in the Poisson algebra $\mathcal{F}(\mathcal{M})$.

The classical BRST charge $Q\in \mathcal{F}(\mathcal{M})$ is
defined as an odd function of ghost number 1 obeying the
\textit{classical master equation}
\begin{equation}\label{nilcon}
    \{Q,Q\}=0\,,
\end{equation}
and the standard boundary conditions
\begin{equation}\label{}
Q|_{\mathcal{P}=0}=\mathcal{C}^aT_a\,.
\end{equation}
A function $a\in \mathcal{F}(\mathcal{M})$ is said to be BRST
invariant if
\begin{equation}\label{}
 DA:=\{Q,A\}=0\,,\qquad {\rm gh}(A)=0\,.
\end{equation}
Clearly, $D^2=0$. The space of \textit{physical observables} is
identified with the zero-ghost-number cohomology of the BRST
operator $D$. The Poisson algebra structure on
$\mathcal{F}(\mathcal{M})$ induces that on the space of physical
observables.

According to general theorems of the BRST theory \cite{HT}, (i)
Eq. (\ref{nilcon}) is always soluble, and (ii) the Poisson algebra
of physical observables is isomorphic to that obtained by the
Hamiltonian reduction by the first class constraints. In the case
at hands, these statements can be refined as follows.

\vspace{5mm}

\noindent \textbf{Proposition 4.1.} \textit{The classical master
equation (\ref{nilcon}) admits the following solution:}
\begin{equation}\label{QQ}
Q=\mathcal{C}^aT_a\,.
\end{equation}
\textit{The Poisson algebra of physical observables on
$\mathcal{M}$ is isomorphic to that on the quasi-symplectic
manifold $(M,\alpha)$. Each physical observable can be represented
by a BRST invariant element from $\mathcal{F}(\mathcal{M})$ that
does not depend on the ghost variables.}

\vspace{5mm}
 \noindent {\bf Proof.}
The first part of the proposition is easily verified by
straightforward calculations. Notice that, unlike what one has in
the standard BRST theory, the first class constraints $T_a$ are no
longer in involution as we have modified the Poisson brackets of
$p$'s by the ghost-dependent term (\ref{modif}). Luckily, this
term does not contribute to the nilpotency condition due to the
symmetry properties of the curvature tensor (\ref{sym}).

The rest of the proposition will follow from the classical limit
of the analogous statement for the quantum BRST observables to be
considered in the next section. Here we just show that each
physical observable  $A(x,p,y,\mathcal{C},\mathcal{P})$ on
$\mathcal{M}$ is uniquely determined by its projection
$A(x,0,0,0,0)$ on $M$. Speaking informally, this implies that the
space of physical observables is not larger than $C^{\infty}(M)$.
In order to see this, let us introduce the following
\textit{homotopy operator} $h: \mathcal{F}(\mathcal{M})\rightarrow
\mathcal{F}(\mathcal{M})$:
\begin{equation}\label{}
h=\mathcal{P}_a\frac{\partial}{\partial \bar p_a}+
y^a\frac{\partial}{\partial \mathcal{C}^a}\,,
\end{equation}
where $\bar{p}_a:= p_a-\omega_{ab}y^b$. From the explicit
expression for the BRST charge (\ref{QQ}) it follows that
\begin{equation}\label{}
    D=\bar p_a\frac{\partial}{\partial \mathcal{P}_a}+
    \mathcal{C}^a\frac{\partial}{\partial y^a}+\cdots\,,
\end{equation}
Here the dots stand for the terms that increase the total degree
when acting on monomials in $y,\mathcal{C},\mathcal{P}$ and $p$.
Then
\begin{equation}\label{}
Dh+hD=N \,,
\end{equation}
where $N=N_0+\cdots$, and
\begin{equation}\label{}
N_0=y^a\frac{\partial}{\partial y^a}+\bar
p_a\frac{\partial}{\partial\bar
p_a}+\mathcal{C}^a\frac{\partial}{\partial \mathcal{C}^a}+
\mathcal{P}_a\frac{\partial}{\partial\mathcal{P}_a}\,.
\end{equation}
Obviously, ${\rm ker}N_0 =C^{\infty}(M)\subset
\mathcal{F}(\mathcal{M})$, and hence the operator $N_0$ is
invertible on the subspace
$\mathcal{F}_0=\mathcal{F}(\mathcal{M})\backslash\, C^{\infty}(M)$
and so is the operator $N$. This implies that the  BRST cohomology
is centered in the subspace $C^{\infty}(M)$; for any BRST
invariant $B$ from the complementary subspace $\mathcal{F}_0$ we
have
\begin{equation}\label{}
B=DC\,,\;\;\;\;\;\;C=h(N|_{\mathcal{F}_0})^{-1}B\,.
\end{equation}

To conclude this section, let us depict the diagram of maps
describing the path from the original quasi-symplectic manifold
$(M,\alpha)$ to the super-Poisson manifold $\mathcal{M}$:
\begin{equation}\label{}
    \mathcal{M}=\mathcal{N}\oplus\Pi\mathcal{N}{\rightarrow}
\mathcal{N}=\mathcal{E}\oplus\mathcal{E}^\ast{\rightarrow}
\mathcal{E}{\rightarrow}M\,.
\end{equation}
All the arrows are canonical projections.

\subsection{Quantization of the super-Poisson manifold
$\mathcal{M}$.} In the general case, it is not easy to quantize
the irregular Poisson brackets (\ref{br3}), (\ref{ghbr}) -
(\ref{modif}). Fortunately, for our purposes, it is sufficient to
deal with a special Poisson subalgebra
$\mathcal{A}\subset\mathcal{F}(\mathcal{M})$. This is given by
functions of $x^\mu, y^a, \mathcal{C}^a$ and
$\mathrm{p}:=\mathcal{C}^ap_a$. Since $\mathrm{p}^2=0$, the
generic element of $\mathcal{A}$ has the form
\begin{equation}\label{}
    A(x,y, \mathrm{p},\mathcal{C})= a(x,y,\mathcal{C})+b(x,y,\mathcal{C})\mathrm{p}\,,
\end{equation}
where $a,b$ belong to the Poisson subalgebra $\mathcal{A}_0$ of
$\mathrm{p}$-independent elements of $\mathcal{A}$. The elements
of $\mathcal{A}_0$ are given thus by formal power series in $y$'s
and $\mathcal{C}$'s with coefficients in $\Lambda (\mathcal{E},
S(\mathcal{E}))$, $S(\mathcal{E})$ being the space of symmetric
tensor powers of $\mathcal{E}$. With this geometrical
interpretation the basic Poisson bracket relations in
$\mathcal{A}$  can be written as
\begin{equation}\label{}
\{\mathrm{p},
\mathrm{p}\}=R+\omega\,,\;\;\;\;\{\mathrm{p},a\}=\nabla
a\,,\;\;\;\;\{a,b\}=\omega^{cd}\frac{\partial a}{\partial
y^c}\frac{\partial b}{\partial y^d}\,, \;\;\;\; a,b \in
\mathcal{A}_0\,.
\end{equation}
Here
\begin{equation}\label{}
\nabla a =\mathcal{C}^d\nabla_d
a=\mathcal{C}^a\left(X^\mu_a\frac{\partial}{\partial
x^\mu}-y^b\Gamma^d_{ab}\frac{\partial}{\partial
y^d}-\mathcal{C}^b\Gamma^d_{ab}\frac{\partial}{\partial
\mathcal{C}^d} \right) a
\end{equation}
is the covariant derivative in $\mathcal{A}_0$ induced by the
symplectic  connection $\nabla_\rho$ on $\Lambda (\mathcal{E},
S(\mathcal{E}))$, and
\begin{equation}\label{}
R=-\frac12R_{abcd}\mathcal{C}^a\mathcal{C}^by^cy^d\,,\;\;\;\;
\omega=\omega_{ab}\mathcal{C}^a\mathcal{C}^b\,,\;\;\;\;\;R,\omega
\in \mathcal{A}_0\,,
\end{equation}
are the covariant curvature tensor and the symplectic form written
in the frame $(y^a, \mathcal{C}^a)$.

In view of Proposition 4.1, the algebra $\mathcal{A}$ contains the
classical BRST charge (\ref{Q}) as well as all the physical
observables of the effective gauge system. It is the reason why
one can restrict consideration to the subalgebra $\mathcal{A}$
when the goal is to quantize the algebra of physical observables.

Proceeding to quantization, we introduce the formal deformation
parameter $\hbar$ and extend the Poisson algebra $\mathcal{A}$ to
the tensor product
\begin{equation}\label{}
    \hat{\mathcal{A}}=\mathcal{A}\otimes [[\hbar]]\,,
\end{equation}
where $[[\hbar]]$ denotes the space of formal power series in
$\hbar$ with coefficients in $\mathbb{C}$. Accordingly, denote by
$\hat{\mathcal{A}}_0:=\mathcal{A}_0\otimes [[\hbar]]$ the
subalgebra of $\mathrm{p}$-independent elements of
$\hat{\mathcal{A}}$. There is an almost obvious quantum product
giving rise to deformation quantization of the Poisson algebra
$\hat{\mathcal{A}}$. For any two elements $a,b\in
\hat{\mathcal{A}}_0$ we just use the Weyl-Moyal formula
\begin{equation}\label{}
    (a\circ b)(x,y,\mathcal{C},\hbar) =\exp\left(-\frac{i\hbar}{2}
    \omega^{ab}\frac{\partial}{\partial y^a}\frac{\partial}{\partial z^b}\right)
    a(x,y,\mathcal{C},\hbar)b(x,z,\mathcal{C},\hbar)|_{y=z}\,.
\end{equation}
and then extend this $\circ$-product to the whole algebra
$\hat{\mathcal{A}}$ by associativity setting
\begin{equation}\label{}
\mathrm{p}\circ a=\mathrm{p}a-i\hbar \nabla
a\,,\;\;\;\;\;a\circ\mathrm{p}=a\mathrm{p}\,,\;\;\;\;\;\mathrm{p}\circ
\mathrm{p}=-i\hbar (R+\omega)\,.
\end{equation}
Clearly, the $\circ$-product respects both the Grassman and the
ghost-number gradings.

As for any graded associative algebra, we can endow
$\hat{\mathcal{A}}$ with the structure of super-Lie algebra w.r.t.
the super-commutator
\begin{equation}\label{}
    [A,B]=A\circ B - (-1)^{\epsilon(A)\epsilon (B)}B\circ
    A\,,
\end{equation}
$A,B$ being homogeneous elements of $\hat{\mathcal{A}}$.

For further purposes let us introduce one more useful grading on
$\hat{\mathcal{A}}$ by prescribing the following degrees to the
variables:
\begin{equation}\label{hdeg}
\deg (x^\mu)=\deg (\mathcal{C}^a)=0\,,\;\;\;\;
\deg(y^a)=1\,,\;\;\;\;\deg (\mathrm{p})=\deg(\hbar)=2\,.
\end{equation}
Since this grading involves essentially the deformation parameter
we will refer to it as \textit{$\hbar$-grading}.

\subsection{Quantum BRST charge.} This is defined as an element
 $\hat{Q}\subset \hat{\mathcal{A}}$ of ghost number 1 satisfying
 the \textit{quantum master equation}
 \begin{equation}\label{meq}
    [\hat{Q},\hat{Q}]=2\hat{Q}\circ \hat{Q}=0
\end{equation}
with the boundary condition
\begin{equation}\label{}
    \hat{Q}|_{\mathcal{P}=\hbar=0}= \mathcal{C}^aT_a\,.
\end{equation}
The adjoint action of $\hat{Q}$ defines the nilpotent derivation
$\hat{D} : \hat{\mathcal{A}}\rightarrow\hat{\mathcal{A}}$:
\begin{equation}\label{}
    \hat{D}a=\frac i\hbar [\hat{Q}, a]\,,\qquad a\in
    \hat{\mathcal{A}}\, .
\end{equation}
The operator $\hat{D}$ increases the ghost number by 1 preserving
the subalgebra $\hat{\mathcal{A}}_0$.

By definition, the space of \textit{quantum physical observables}
is identified with the zero-ghost-number cohomology of the
operator $\hat{D}$.

Let us show the existence of a quantum BRST charge $\hat{Q}$ whose
classical limit coincides with the classical BRST charge $Q$.
Technically, instead of finding $\hbar$-corrections to $Q$, it is
more convenient to build up $\hat{Q}$ using  recursion on the
total $\hbar$-degree (\ref{hdeg}). In order to do this we
introduce the pair of Fedosov's operators changing the
$\hbar$-degree by 1 unit.  The first operator is given by
\begin{equation}\label{}
\delta a = \mathcal{C}^a\frac{\partial a}{\partial
y^a}\,,\;\;\;\;\;\delta^2=0\,.
\end{equation}
for any $a\in \hat{\mathcal{A}}_0$.  Since
\begin{equation}\label{}
    \delta a= \frac {i}{\hbar}[\mathcal{C}^a\omega_{ab}y^b, a]\,,
\end{equation}
it is an internal derivation of $\hat{A}_0$. The second operator
is defined by its action on homogeneous functions:
\begin{equation}\label{}
\begin{array}{l}
   \displaystyle \delta^\ast a_{mn}=\frac{1}{n+m}y^a\frac{\partial a}{\partial
\mathcal{C}^a}\,,\;\;\;\;\;\;n+m\neq 0\,,\\[3mm]
   \delta^{\ast}a_{00}=0\,,
\end{array}
\end{equation}
where $a_{nm}=a_{a_1\cdots a_n b_1\cdots
b_m}(x,\hbar)y^{a_1}\cdots y^{a_n}\mathcal{C}^{b_1}\cdots
\mathcal{C}^{b_m}$. Like $\delta$, the operator $\delta^\ast$ is
nilpotent, though it is not a derivation of the $\circ$-product.
One can regard $\delta^\ast$ as a homotopy operator for $\delta$:
\begin{equation}\label{HdR}
    a|_{\mathcal{C}=y=0}+ \delta \delta^\ast a+\delta^*\delta a = a\,,\;\;\;\;\;
    \forall a\in \hat{\mathcal{A}}_0\,.
\end{equation}
The last relation resembles the usual Hodge-De Rham decomposition
for the exterior algebra of differential forms.

\vspace{5mm}\noindent {\bf Proposition 4.2.} {\it The quantum
master equation (\ref{meq}) has a solution of the form
\begin{equation}\label{qq}
\hat{Q}=\sum_{r=1}^\infty Q_{r}\,,\;\;\;\;\;\;\;\deg (Q_r) =r\,,
\end{equation}
where
\begin{equation}\label{}
  Q_1=-\mathcal{C}^a\omega_{ab}y^b\,,\;\;\;\;\;\;\;Q_2\,,
  =\mathrm{p}\,,\quad {and} \quad Q_{r}\in \hat{\mathcal{A}}_0\,,\;\forall
  r>2\,,
  \end{equation}
which is unique if we require
  \begin{equation}\label{dQ}
     \delta^*Q_{r}=0\,,\;\;\;\; \forall r>2\,.
\end{equation}}

\vspace{5mm}\noindent {\bf Proof.} The first three terms in the
expansion (\ref{qq}) coincide with those in the classical BRST
charge (\ref{QQ}), and this proves the validity of Eq. (\ref{meq})
in the lowest order in $\hbar$-degree. For $r\geq 4$ Eq.
(\ref{meq}) implies
\begin{equation}\label{QB}
    \delta Q_{r+1}=B_r\,,
\end{equation}
where
\begin{equation}\label{}
    B_r=-\frac{i}{2\hbar}\sum_{s=0}^{r-2} [Q_{r+2},Q_{r-s}]\,,\;\;\;\;\;\deg (B_r)=r\,.
\end{equation}
In view of the Hodge-De Rham decomposition (\ref{HdR}),  Eq.
(\ref{QB}) is soluble iff  $\delta B_r=0$. In this case we have
the unique solution
\begin{equation}\label{Qrec}
Q_{r+1}=\delta^*B_r
\end{equation}
subject to the extra condition $\delta^*Q_{r+1}=0$. So, it remains
to show that $\delta B_r=0$. Proceeding by induction, we assume
that Eq. (\ref{meq}) is valid up to the $s$-th order in
$\hbar$-degree. Then, extracting the $(s+3)$-order in  the Jacobi
identity
\begin{equation}\label{}
    [Q^s,[Q^s,Q^s]]=0\,,\;\;\;\;\;\;Q^s=\sum_{k=1}^sQ_k\,,
\end{equation}
one gets $\delta B_s=0$, that completes the proof.

\vspace{5mm} Since $\mathrm{gh}(Q)=1$, the ansatz (\ref{qq})
implies the following structure of the quantum BRST charge:
$Q=\mathcal{C}^a\hat{T}_a(x,y,\hbar)$, where $\hat{T}_a$ are the
``quantum" first class constraints. Then Rel. (\ref{dQ}) is just
another form of the $y$-transversality condition (\ref{yT}), which
allows one to extract a unique solution both at the classical and
quantum levels.

\subsection{Quantum observables and star-product.}
In Sect. $4.1.$  we have shown  that the space of physical
observables of the classical gauge system on $\mathcal{N}$ is not
larger than $C^{\infty}(M)$ in the sense that any physical
observable is uniquely determined by its projection on $M$. In
this section we prove the inverse: any physical observable on $M$,
has a unique BRST-invariant extension to a zero-ghost-number
function from $\mathcal{A}_0$. Moreover, this picture takes place
at the quantum level as well if we replace $C^\infty(M)\rightarrow
C^{\infty}(M)\otimes [[\hbar]]$ and $\mathcal{A}_0 \rightarrow
\hat{\mathcal{A}}_0$. Therefore, it is sufficient to consider only
the quantum case, the classical statement will follow from the
classical limit.

\vspace{5mm}\noindent{\bf Proposition 4.3.} {\it For any $a\in
C^\infty(M)$ there is a unique $\hat{a}\in \hat{\mathcal{A}}_0$
obeying conditions }
\begin{equation}\label{mst}
[\hat{Q},\hat{a}]=0\,,\;\;\;\;\;\hat{a}|_{y=0}=a \,.
\end{equation}

\vspace{5mm}\noindent{\bf Proof.} Consider the expansion of
$\hat{a}\in \hat{\mathcal{A}}_0$ according to the $\hbar$-degree:
\begin{equation}\label{}
\hat{a}=\sum_{s=0}^\infty
{a}_s\,,\;\;\;\;\;\;\;\;\;\;\deg({a}_s)=s\,.
\end{equation}
The second condition in (\ref{mst}) says that $a_0=a(x)$.
Substituting this expansion into the first equation, one gets
\begin{equation}\label{dab}
    \delta a_{s+1}=B_s\,,\;\;\;\;\;\;s>1\,,
\end{equation}
where
\begin{equation}\label{}
    B^s=\frac i\hbar\sum_{k=0}^{s-2}[Q^{k+2},a^{s-k}]\,,\;\;\;\;\;\;\;
    \deg(B^s)=s\,.
\end{equation}
In view of the Hodge-De Rham decomposition (\ref{HdR}) and the
boundary condition (\ref{mst}), Eq. (\ref{dab}) has the unique
solution
\begin{equation}\label{arec}
    a^{s+1}=\delta^*B^s\,,
\end{equation}
provided the r.h.s. is $\delta$-closed. The equality $\delta
B^s=0$ follows by induction  from the $(s+3)$-order of the
identity
\begin{equation}\label{}
    [\hat{Q},[\hat{Q},\hat{a}]]=0\,.
\end{equation}

\vspace{5mm} \noindent {\bf Corollary.} {\it There is a linear
isomorphism between the spaces of quantum observables on $M$,
i.e., $C^{\infty}(M)\otimes [[\hbar]]$, and the zero-ghost-number
cohomology of the BRST-differential $\hat{D}: \hat{\mathcal{A}}_0
\rightarrow \hat{\mathcal{A}}_0$.}

\vspace{5mm}\noindent{\bf Proof.} Eq. (\ref{mst}), being linear,
has a unique solution  even though we allow $\hat{a}|_{y=0}$ to
depend formally on $\hbar$. Therefore, we can replace
$C^\infty(M)$ with $C^{\infty}(M)\otimes [[\hbar]]$.

\vspace{5mm} Clearly, the $\circ$-product on $\hat{\mathcal{A}}_0$
descends to the BRST cohomology and, in view of the corollary,
induces an associative $\ast$-product on $C^\infty(M)\otimes
[[\hbar]]$. Explicitly,
\begin{equation}\label{st-pr}
    a\ast b =(\hat{a}\circ
    \hat{b})|_{y=0}=\sum_{n=1}^\infty\hbar^nD_n(a,b)
    \,,\qquad
    \forall a,b\in C^{\infty}(M)\otimes [[\hbar]]\,,
\end{equation}
where
\begin{equation}\label{}
    D_0(a,b)=ab\,,\qquad D_1(a,b)=-\frac{i}2 \{a,b\}_{M}\,,
\end{equation}
and the ``hat" stands for the BRST-invariant lift from $M$ to
$\mathcal{M}$  (the existence and uniqueness of such a lift are
ensured by Proposition 4.3). The higher orders in $\hbar$, being
recurrently constructed by (\ref{Qrec}) and (\ref{arec}), involve
also the symplectic connection and the curvature.

\vspace{5mm}\noindent{\textit{Remark}.} By construction, the
bi-differential operators $D_n$ entering the $\ast$-product
(\ref{st-pr}) have a rather peculiar structure. Namely, they are
determined  by repeated differentiations along the anchor
distribution $\{X_a\}$:
\begin{equation}\label{E-dif}
    D_n(a,b)=\sum_{k,l\leq n}D_n^{c_1\cdots c_k d_1\cdots
    d_l}(x)(X_{c_1}\cdots X_{c_k}a)(X_{d_1}\cdots X_{d_l}b)\,.
\end{equation}
Here the structure functions $D^{c_1\cdots c_k d_1\cdots d_l}(x)$
are \textit{universally} expressed via the data of a symplectic
Lie algebroid and a Lie algebroid connection. The differential
operators of the form (\ref{E-dif}) are called the
$\mathcal{E}$-differential operators; accordingly, the
$\ast$-product (\ref{st-pr}) is called the
$\mathcal{E}$-deformation of $M$. As was shown in \cite{NT}, any
$\mathcal{E}$-deformation of $M$ can be induced by an
$\mathcal{E}$-deformation of  the commutative algebra of
$\mathcal{E}$-jets. Conversely, the $\mathcal{E}$-deformation of
$M$, given by the formula (\ref{st-pr}), admits a canonical
extension to the space of $\mathcal{E}$-jets (by universality). In
\cite{DILSh} such an extension was used to derive the universal
deformation formula for triangular Lie bialgebras.

\section{Factorizable Poisson brackets beyond symplectic Lie algebroids.}

As we have seen, the concept of a symplectic Lie algebroid gives
rise to an interesting class of Poisson brackets. Not any Poisson
bracket comes in this way, but when it does, we have a simple
quantization procedure generalizing the Fedosov quantization. In
this section we would like to discuss, in a sense, an inverse
problem: To what extent the factorization (\ref{biv}) of a Poisson
bi-vector $\alpha$ defines a symplectic Lie algebroid?

The precise formulation of the problem is as follows.   Let
$\mathcal{E}\rightarrow M$ be a vector bundle over a smooth
manifold $M$, $\omega$ a section of
$\mathcal{E}\wedge\mathcal{E}$, and $X$ a section of
$\mathcal{E}^\ast\otimes TM$. By a slight abuse of notation, we
will use the same letters $\omega$ and $X$ to denote the
corresponding bundle homomorphisms $\omega :
\mathcal{E}^\ast\rightarrow \mathcal{E}$ and $X:
\mathcal{E}\rightarrow TM$. Let us also suppose that the
$\mathcal{E}$-bi-vector $\omega$ is non-degenerate (defines an
isomorphism between $\mathcal{E}$ and its dual $\mathcal{E}^\ast$)
and $X$ is involutive. The latter means that in each trivializing
coordinate chart $\mathcal{U}\subset M$ with frame $s_\alpha \in
\Gamma(\mathcal{E}|_{\mathcal{U}})$, the local vector fields
$X_\alpha=X_\alpha^i \partial_i \in {\rm Vect}(\mathcal{U})$ form
an involutive distribution,
\begin{equation}\label{involution}
    [X_\alpha,X_\beta]=f_{\alpha\beta}^\gamma X_\gamma\,,
\end{equation}
$f_{\alpha\beta}^\gamma$ being smooth functions on $\mathcal{U}$.
Clearly, the property of $\{ X_\alpha \}$ to be involutive does
not depend on a frame, and hence, $\{X_a\}$ generates a (singular)
foliation $F(M)$. Suppose now that the bi-vector
\begin{equation}\label{dalph}
    \alpha=\omega^{\alpha\beta}X_\alpha\wedge
    X_\beta \in \wedge ^2 TM
\end{equation}
satisfies the Jacobi identity
\begin{equation}\label{alpha}
    [\alpha,\alpha]=0\,.
\end{equation}

\noindent {\bf Question:} \textit {What is the most general
geometric structure underlying  Eqs.
(\ref{involution}-\ref{alpha}). }

\vspace{3mm} A particular solution to these equations is delivered
by a symplectic Lie algebroid $\mathcal{E}\rightarrow M$ with
anchor $X$ and symplectic 2-form $\omega$. In this case
$(M,\alpha)$ is just a quasi-symplectic manifold considered in the
previous sections.

Explicitly, the Jacobi identities for the local vector fields
$X_a$ and the Poisson bi-vector $\alpha$ read
\begin{equation}\label{sse}
    (f_{\alpha\beta}^\delta f_{\delta\gamma}^\mu-X_\gamma f_{\alpha\beta}^\mu+
    cycle(\alpha,\beta,\gamma))X_\mu=0\,,
\end{equation}
\begin{equation}\label{sse1}
\begin{array}{c}
 (\omega_{\gamma\delta}f_{\alpha\beta}^\delta -X_\gamma
 \omega_{\alpha\beta})X^\alpha\wedge X^\beta \wedge
 X^\gamma=0\,,\\[3mm]
 X^\alpha=\omega^{\alpha\beta}X_\beta\,,\;\;\;\;\;
 \omega_{\alpha\gamma}\omega^{\gamma\beta}=\delta_\alpha^\beta\,.
 \end{array}
\end{equation}
If the map $X: \mathcal{E}\rightarrow TM$ is injective on an
everywhere dense domain in $M$, the expressions in  parentheses
(\ref{sse1}) must vanish by continuity, and we arrive at a
symplectic Lie algebroid $(\mathcal{E},X, \omega)$. In the
opposite case the l.h.s. of Eqs. (\ref{sse}), (\ref{sse1}) cannot
be ``divided'' by $X_a$ so simply, and thus, more general
solutions for the structure functions $f_{\alpha\beta}^\gamma$,
$\omega_{\alpha\beta}$ and $X_\alpha^i$ are possible. To further
study these equations, we impose a certain regularity condition on
$X$. In what follows we will assume that the space $\Gamma(TM)$,
considered as a $C^\infty(M)$-module, admits a resolution of the
form
\begin{equation}\label{res}
    0\leftarrow TM \stackrel{\;\;\;d_1}{\leftarrow}
    \mathcal{E}_{1}\stackrel{\;\;\;d_2}{\leftarrow}
    \mathcal{E}_2\leftarrow \cdots {\leftarrow}\; \mathcal{E}_{n-1}
    \stackrel{\;\;\;d_n}{\leftarrow}\mathcal{E}_n \leftarrow 0\,,
\end{equation}
where $\mathcal{E}_k\rightarrow M$ and $d_k$ are sequences of
vector bundles over $M$ and their $M$-morphisms with
$\mathcal{E}_1=\mathcal{E}$ and $d_1=X$. (A sequence of
homomorphisms of modules (\ref{res}) is called a resolution of the
module $\Gamma(TM)$, if ${\rm im}\, d_{k+1}=\ker d_k$. In other
words, the sequence (\ref{res}) is just a cochain complex, which
is exact, exclusive of maybe the first term.) Here we do not
require the morphisms $d_k$ to have constant ranks, but since $n<
\infty$, their ranks have to be constant on an open everywhere
dense domain in $M$. In particular, the last structure map
$d_k:\mathcal{E}_k\rightarrow \mathcal{E}_{k-1}$ should be
injective on an everywhere dense domain in $M$. By analogy with
ordinary Lie algebroids, we will refer to the first structure map
$d_1=X$ as anchor.

 In order to clarify the meaning of the regularity condition,
 let us choose an open domain $\mathcal{U}\subset M$ such that for
 all $k=1,...,n$, $\mathcal{E}_k|_\mathcal{U}$ is a trivial vector
bundle with frame $s_{\alpha_k}$. Upon restriction on
$\mathcal{U}$, the morphisms $d_k$ are represented by matrices
$d^{\alpha_{k-1}}_{\alpha_k}$, so that
$d_{\alpha_k}^{\alpha_{k-1}}d_{\alpha_{k+1}}^{\alpha_k}=0$. Since
the complex (\ref{res}) is exact starting with $d_1$, the equality
$f^{\alpha_k}d_{\alpha_k}^{\alpha_{k-1}}=0$, $f^{\alpha_k}$ being
a section of $\mathcal{E}_k$, implies that
$f^{\alpha_{k}}=g^{\alpha_{k+1}}d^{\alpha_k}_{\alpha_{k+1}}$ for
some section  $g^{\alpha_{k+1}}$ of $\mathcal{E}_{k+1}$.

\vspace{5mm} \noindent  {\it Example 0.} Consider the adjoint
representation of $so(3)$. Identifying the carrier space $so(3)$
with $\mathbb{R}^3$ we get a set of three linear vector fields on
$\mathbb{R}^3$ generating the $so(3)$-algebra action:
    \begin{equation}\label{}
    \mathrm{ad}_i=\epsilon_{ijk}x_j\partial_k
    \,,\;\;\;\;\;\;\;\;\;[\mathrm{ad}_i,\mathrm{ad}_j]=
    -\epsilon_{ijk}\mathrm{ad}_k\,.
\end{equation}
Clearly, the rank of the anchor $\mathrm{ad} : \mathbb{R}^3\times
so(3) \rightarrow T\mathbb{R}^3$ equals $2$ in general position
and vanishes at the origin $0\in \mathbb{R}^3$. Since  the
equation $f^i(x) \mathrm{ad}_{i}=0$ implies $f^i(x)=g(x)x^i$, for
some smooth function $g$, while the equation $x^ih(x)=0$ has the
unique solution $h=0$, we get the following resolution:
\begin{equation}\label{}
    0\leftarrow T\mathbb{R}^3\stackrel{\rm \;\;ad}{\leftarrow}\mathbb{R}^3
    \times so(3)\stackrel{\;\;d_2}{\leftarrow}
    \mathcal{E}_2 \leftarrow 0\,,
\end{equation}
were $\mathcal{E}_2$ is a linear bundle over $\mathbb{R}^3$ and
$d_2=(x^i)$.

\vspace{5mm}

Given an anchor $X$ satisfying the regularity condition, one can
solve the Jacobi identity (\ref{sse}) in the following form:
\begin{equation}\label{se}
    f_{\alpha\beta}^\delta f_{\delta\gamma}^\mu-X_\gamma f_{\alpha\beta}^\mu+
    cycle(\alpha,\beta,\gamma)=f_{\alpha\beta\gamma}^ad_a^\mu \,,
\end{equation}
where $f_{\alpha\beta\gamma}^a$ are smooth functions on $
\mathcal{\mathcal{U}}$, skew-symmetric in $\alpha\beta\gamma$, and
$d^\alpha _a$ is the matrix of the second structure map $d_2$ in
(\ref{res}). By definition, we have
\begin{equation}\label{dX}
d_a^\gamma X_\gamma=0\,.
\end{equation}
In order to solve the Jacobi identity for $\alpha$, we assume the
anchor foliation $F(M)$ to be regular\footnote{This technical
restriction can be weakened.}, i.e., ${\rm im} X$ is an integrable
subbundle of $TM$. Then
\begin{equation}\label{se1}
 \omega_{\gamma\delta}f_{\alpha\beta}^\delta -X_\gamma
 \omega_{\alpha\beta}+cycle(\alpha,\beta,\gamma)=
 W_{\alpha\beta}^ad_{a\gamma}+cycle(\alpha,\beta,\gamma) \,,\;\;\;\;\;\;
 d_{a\gamma}=\omega_{\gamma \beta}d^\beta_a\,,
\end{equation}
$W_{\alpha\beta}^a=-W_{\beta\alpha}^a$ being smooth functions on
$\mathcal{U}$. Examining compatibility of these equations with the
involution relations (\ref{involution}), one obtains an infinite
set of higher structure functions and structure relations to be
studied below.

Let us forget for a moment about the Poisson bi-vector $\alpha$,
focusing at the anchor distribution $X$.  Commuting (\ref{dX})
with $X_\beta$, we find
\begin{equation}\label{}
X_\beta d^\alpha_a+ f_{\beta\gamma}^\alpha d^\gamma_a = -f_{\beta
a}^b d_{b}^\alpha\,,
\end{equation}
$f_{b a}^\beta$ being local functions.  Contracting the last
identity with $d^\beta_c$ and symmetrizing  in the indices $a c$,
we get
\begin{equation}\label{}
d^\beta_c f_{\beta a}^b + d^\beta_a f_{\beta c}^b=f_{c
a}^Ad_A^b\,,
\end{equation}
where $d_A^b$ is the matrix of the third structure map in
(\ref{res}), so that $d_A^b d_b^\alpha=0$. Proceeding in the same
manner one can derive the other structure relations.

There is a nice way to generate all these relations systematically
using the language  of \textit{NQ-manifolds}.  Let us recall the
basic definitions \cite{Vain}, \cite{Sev}, \cite{Vor}, \cite{Roy}.
An N-\textit{manifold} is a non-negatively integer graded
supermanifold, whose N-grading is compatible with the underlying
$\mathbb{Z}_2$-grading (Grassman parity). In other words, an
N-manifold is just a supermanifold with a privileged class of
atlases in which particular coordinates are assigned non-negative
integer degrees (even coordinates have even degrees, while odd
ones have odd degrees) so that the changes of coordinates respect
these degrees. The highest degree of coordinates is called the
\textit{degree } of an N-manifold. For example, if $\deg M=0$,
then $M$ is just an ordinary manifold. Finally, an NQ-manifold is
an N-manifold endowed with integrable vector field $Q$ of degree
$1$, called a \textit{homological vector field}. Since $Q$ is odd,
the integrability condition $[Q,Q]=2Q^2=0$ is nontrivial. The
classical example of an NQ-manifold of degree $1$ is the
anti-cotangent bundle $\Pi TM$ (cotangent bundle with the reverse
parity of fibers). The functions on $\Pi TM$ are just
inhomogeneous differential forms on $M$ and $Q$ is the usual
exterior differential. More generally, an NQ-manifold of degree 1
is the same as  Lie algebroid. For this reason it is natural to
name the NQ-manifolds of degree $n$ as $n$-{\it algebroids}
\cite{Roy}.

A general homological vector field looks like\footnote{All
derivatives are assumed to be acting on the left.}
\begin{equation}
    Q=c^{\alpha_1}X_{\alpha_1}^i(x)\frac\partial{\partial x^i}+\sum _{k=1}^{\deg
    \mathcal{M}}c^{\alpha_{k+1}}d_{\alpha_{k+1}}^{\alpha_k}(x)
    \frac{\partial}{\partial c^{\alpha_k}}+\cdots \,,
\end{equation}
where $\deg (x^i)=0$, $\deg (c^{\alpha_k})=k$, and dots stand for
higher orders in the positively graded variables $c^{\alpha_k}$.
Evaluating the equation $Q^2=0$ at the first order in $c$'s one
recovers the cochain complex axioms
$d_{\alpha_m}^{\alpha_{m-1}}d_{\alpha_{m+1}}^{\alpha_m}=0$,  the
second order in $c^{\alpha_1}$ reproduces the involution relations
(\ref{involution}), Rel. (\ref{se}) contributes to the cubic
order, and so on \footnote{Notice, that any $n$-algebroid can also
be viewed as an $(n+1)$-algebroid whose higher structure functions
just equal to zero.}. Thus, we see that the resolution (\ref{res})
for the involutive distribution $X: \mathcal{E}\rightarrow TM$ is
just a regular $n$-algebroid.

Although the language of NQ-manifolds is quite convenient to
describe the structure of $n$-algebroids as such, it becomes
unappropriate when one tries to incorporate the symplectic
structure entering the factorization (\ref{biv}). Here we would
like to present a new geometric framework  providing a uniform
description for both $n$-algebroid and  symplectic structures
underlying factorizable Poisson brackets. For the sake of
simplicity we restrict our consideration to the case of
2-algebroids. The general construction  will be developed
elsewhere. Before going into details let us give two examples
which are of interest by themselves.

\vspace{5mm} \noindent{\it Example 1.} (Poisson-Lie algebras.)
Consider an invariant Poisson bracket on a Lie group $G$
associated with the bi-vector
\begin{equation}\label{}
\alpha =r^{ij}(L_i\wedge L_j-R_i\wedge R_j)\,.
\end{equation}
Here $L_i$, $R_j$ are left and right invariant vector fields on
$G$, and the matrix $(r^{ij})$ obeys the Yang-Baxter equation
\begin{equation}\label{}
    f^i_{ml}\left(r^{jn}f_{ns}^l r^{sk}
    +cycle(j,l,k)\right)+cycle(i,j,k)=0\,,
\end{equation}
$f_{ij}^k$ being the structure constants of the corresponding Lie
algebra $L(G)$. If $\det (r^{ij})\neq 0$, we have the Poisson
bi-vector $\alpha$ associated with the symplectic structure
$r^{ij}$ and the 2-algebroid
\begin{equation}\label{}
0\leftarrow TG\stackrel{\;\;\;(L,R)}{\leftarrow} TG\oplus TG
\stackrel{\;\;\;d_2}{\leftarrow}TG\leftarrow 0\,,
\end{equation}
where $d_2=(1,A)$, and $A$ is the automorphism of the tangent
bundle $TG$ relating the left and right invariant vector fields:
$L_i=A_i^jR_j$.

\vspace{5mm}\noindent{\it Example 2.} (Universal factorization.)
Any Poisson bi-vector
$\alpha=\alpha^{ij}\partial_i\wedge\partial_j $ can be factorized
in a skew-symmetric product of  Hamiltonian and coordinate vector
fields:
\begin{equation}\label{}
    \alpha = P_i\wedge
    Q^i\,,\;\;\;\;\;\;\;\;P_i=\partial_i\,,\;\;\;Q^i=\alpha^{ij}\partial_j\,.
\end{equation}
The local vector distribution $(P_j,Q^j)$ is obviously transitive
and hance involutive.  Moreover, there is a one-parameter
ambiguity in writing the involution relations:
\begin{equation}\label{}
    \begin{array}{l}
 [P_i,P_j]=0\,,\\[3mm]
 [P_i,Q^j]=\partial_i\alpha^{jk}P_k\,,\\[3mm]
 [Q^i,Q^j]=t\partial_k\alpha^{ij}Q^k+(1-t)\partial_k\alpha^{ij}\alpha^{kn}P_n\,,\;\;\;\;\;
 t\in \mathbb{R}\,.
    \end{array}
\end{equation}
This ambiguity is due to linear dependence of the local vector
fields:
\begin{equation}\label{}
    Q^i=\alpha^{ij}P_j\,.
\end{equation}
The last equations are already independent and we arrive at the
following cochain complex
\begin{equation}\label{}
    0 \leftarrow TM\stackrel{(P,Q)}{\longleftarrow} TM\oplus T^*M
    \stackrel{(1,\alpha)}{\longleftarrow} TM\leftarrow 0\,,
\end{equation}
which is exact provided $\alpha$ is non-degenerate on an
everywhere dense domain in $M$.  \vspace{5mm}

Consider now a general NQ-manifold $\mathcal{M}$ of degree $2$. As
for usual manifolds, the structure of $\mathcal{M}$ can be
described in terms  of coordinate charts and transition functions
gluing together individual N-graded domains $U\in \mathcal{M}$.
Without loss of generality we can assume that each $U$ is given by
a direct product $\mathcal{U}\times
\mathbb{R}^n_1\times\mathbb{R}^m_2$, where $\mathcal{U}\in M$ is
an open contractible domain on the base manifold with local
coordinates $x^i$, $\mathbb{R}^n_1$ and $\mathbb{R}^m_2$ are
vector spaces with linear coordinates $c^\alpha$ and $c^a$,
respectively. We set $\deg (x^i) =0$, $\deg (c^\alpha)=1$, $\deg
(c^a)=2$, so that $x^i$ and $c^a$ are commuting, while $c^\alpha$
are anticommuting coordinates on $U\in \mathcal{M}$. If now $U$
and $U{}'$ are two graded domains with nonempty intersection, then
the most general form of  transition functions, compatible with
the N-grading, is given by
\begin{equation}\label{trf}
    x'^i=f^i(x)\,,\;\;\;\;\;\;\;\;
    c'^\alpha =A^{\alpha}_{\beta}(x)c^\beta\,,\;\;\;\;\;\;\;
    c'^a=B^a_b(x) c^b+\frac12F_{\alpha\beta}^a(x)c^\alpha c^\beta\,,
\end{equation}
$f^i, A^\alpha_\beta, B^a_b$ being smooth functions on
$\mathcal{U}\cap \mathcal{U}\,{}'$. The first equation defines
transformation of local coordinates on the base manifold $M$.
Disregarding the $F$-term, we see that the second and third
equations are similar to those defining transition functions for
(graded) vector bundles. Moreover, the matrix-valued functions $A$
and $B$ do really obey the standard cocycle conditions on overlaps
of two and three coordinate charts, defining thus direct sum
$\mathcal{E}_1\oplus\mathcal{E}_2$ of two graded vector bundles.

In terms of local coordinates the  most general homological vector
field on $\mathcal{M}$ reads
\begin{equation}\label{Q}
    Q=c^\alpha X_\alpha^i\frac{\partial}{\partial
    x^i}+c^ad_a^\alpha\frac{\partial}{\partial
    c^\alpha}+\frac12 c^\beta c^\alpha f^\gamma_{\alpha \beta}
    \frac{\partial}{\partial c^\gamma}+c^\alpha c^a f_{\alpha
    a}^b\frac{\partial}{\partial c^b}+
    c^\gamma c^\beta c^\alpha
    f_{\alpha \beta \gamma}^a
    \frac{\partial}{\partial
    c^a}\,.
\end{equation}
Using relations (\ref{trf}) one can derive transformation
properties for the structure functions $X_\alpha^i$, $d^\gamma_a$,
$f_{\alpha\beta}^\gamma$, $f_{a\beta}^b$,
$f_{\alpha\beta\gamma}^a\in C^\infty (\mathcal{U})$ under
coordinate changes. In particular, the $F$-term induces the shift
\begin{equation}\label{shift}
f_{\alpha\beta}^\gamma \rightarrow
f_{\alpha\beta}^\gamma+F_{\alpha\beta}^a d_a^\gamma\,,
\end{equation}
which reflects an inherent ambiguity concerning the choice of the
structure functions (\ref{involution}) whenever $X_\alpha$ are
linearly dependent. Also, it is not hard to see that $X^i_\alpha$
and $d_a^\gamma$ transform homogeneously, i.e., as sections of the
associated vector bundles $\mathcal{E}^\ast_1 \oplus TM$ and
$\mathcal{E}^\ast_2\oplus\mathcal{E}_1$.

Now suppose that $\mathcal{M}$ defines a 2-algebroid factorizing a
Poisson bi-vector $\alpha$. Our aim is to give a unified
description for both the 2-algebroid and the symplectic structure
entering this factorization. It turns out that all structure
relations underlying the factorization (\ref{dalph}) can be
described in terms of an abelian connection (covariant derivative)
acting on a bundle of odd Poisson algebras over $M$. The
construction goes as follows.

Let ${E}_0\oplus {E}_1$ be a $\mathbb{Z}_2$-graded vector bundle
over $M$ defined by aforementioned gluing cocycles $A$ and
$(B^{-1})^\ast$, that is $E_1=\mathcal{E}_1$,
$E_0=\mathcal{E}_2^\ast$. If $c^\alpha$ and $\pi_a$ are linear
coordinates in the fibers of ${E}_1$ and ${E}_0$ over a
trivializing domain $\mathcal{U}\in M$, we set
$\epsilon(c^\alpha)=1$, $\epsilon(\pi_a)=\epsilon(x^i)= 0$. It is
convenient to think of this bundle as a formal supermanifold
$\mathcal{N}$ with  even coordinates $x^i, \pi_a$ and odd
coordinates $c^\alpha$.  The word ``formal'' reflects the fact
that we allow the functions on $\mathcal{N}$ to be given by formal
power series in $\pi$'s. These functions  form a supercommutative
algebra  $\mathcal{F}$ with  the generic element
\begin{equation}\label{}
f(x,c,\pi)=\sum_{k,n}f_{\alpha_1\cdots\alpha_n}^{a_1\cdots a_k
}(x)\pi_{a_1}\cdots \pi_{a_k}c^{\alpha_1}\cdots c^{\alpha_n}\,.
\end{equation}
The algebra $\mathcal{F}=\oplus \mathcal{F}_{n,m}$ is naturally
bigraded w.r.t. powers of $c$'s and $\pi$'s and is isomorphic to
the tensor algebra of sections of the associated vector bundle
$S^\bullet {E}^*_0\otimes \Lambda^\bullet {E}_1^\ast$.

The space $\mathcal{F}$ can also be endowed with the structure of
odd Poisson algebra. To this end, we introduce the odd Laplacian
$\Delta: \mathcal{F}_{m,n}\rightarrow \mathcal{F}_{m-1,n-1}$:
\begin{equation}\label{}
    \Delta f=d^\alpha_a(x)\frac{\partial ^2 f}{\partial c^\alpha \partial
    \pi_a}\,.
\end{equation}
Clearly, $\Delta^2=0$. The odd Poisson bracket
$(\,\cdot\,,\,\cdot\,):\mathcal{F}_{n,m}\otimes
\mathcal{F}_{k,l}\rightarrow \mathcal{F}_{n+k-1,m+l-1}$ is defined
by the rule:
\begin{equation}\label{oddbr}
    (-1)^{\epsilon(f)}(f,g)= \Delta (f\cdot g) -\Delta f\cdot g -(-1)^{\epsilon (f)}
    f\cdot \Delta g\,.
\end{equation}
It obeys the standard identities which may be taken as the axioms
of an odd Poisson manifold:
\begin{equation}\label{}
    \begin{array}{ll}
    \epsilon (f,g)=\epsilon(f) +\epsilon(g)+1&  ({\rm {mod}}\,\; 2) \,,\\[3mm]
      (f,g)=-(g,f)(-1)^{(\epsilon (f)+1)(\epsilon(g)+1)}&
      (symmetry)\,, \\[3mm]
      (f,gh)=(f,g)h+(f,h)g (-1)^{\epsilon (g)\epsilon (h)}& (Leibnitz\;\;rule)\,,
      \\[3mm]
       (f,(g,h))(-1)^{(\epsilon (f)+1)(\epsilon (h)+1)}+cycle(f,g,h)=0&
       (Jacobi\;\; identity)\,.
    \end{array}
\end{equation}
Notice that $\Delta$ respects the odd Poisson bracket
(\ref{oddbr}) in the sense that
\begin{equation}\label{}
    \Delta (f,g)=(\Delta f,g)+(-1)^{\epsilon (f)+1}(f,\Delta
    g)\,.
\end{equation}

The algebra $\mathcal{F}$ contains a special element
$\omega=\frac12\omega_{\alpha\beta}c^\alpha c^\beta\in
\mathcal{F}_{2,0}$ associated with the symplectic structure. The
adjoint action of $\omega$ gives rise to the nilpotent
differentiation $\delta : \mathcal{F}_{n,m}\rightarrow
\mathcal{F}_{n+1,m-1}$:
\begin{equation}\label{}
    \delta f= (\omega, f)=-c^\alpha\omega_{\alpha\beta}d^\beta_a
    \frac{\partial f}{\partial \pi_a
    }\,,\qquad \delta^2=0\,.
\end{equation}
It easy to see that the $\delta$-cohomology is trivial when
evaluated on $\mathcal{F}_{\bullet,k}$ with $k>0$ .

Now we would like to endow the bundle of odd Poisson algebras
$\mathcal{F}$ with a sort of partial connection  $\nabla :
\mathcal{F}_{m,n}\rightarrow \mathcal{F}_{m+1,n}$ making possible
parallel transport along the leaves of the anchor foliation
$F(M)$. Treating  $\nabla$ as an odd vector field on
$\mathcal{N}$, we set
\begin{equation}\label{oddvf}
    \nabla a=c^\alpha\left(X_\alpha^i\frac{\partial}{\partial x^i} + \frac12
    c^\beta f_{\alpha\beta}^\gamma\frac{\partial}{\partial c^\gamma}+ \pi_a f^a_{\alpha b}
    \frac{\partial}{\partial \pi_b}\right) a + (W_1, a)\,,
\end{equation}
where the structure functions $X_\alpha^i$,
$f_{\alpha\beta}^\gamma$, $f^a_{\alpha b}$ are the same as in Eq.
(\ref{Q}) and $W_1=c^\alpha c^\beta W_{\alpha\beta}^a\pi_a\in
\mathcal{F}_{2,1}$ is given by the r.h.s. of Eq. (\ref{se1}).
Using the definition  of $\nabla$, one can rewrite Eq. (\ref{se1})
as
\begin{equation}\label{xxx}
\nabla \omega =0
\end{equation}
or, equivalently,
\begin{equation}\label{ndel}
    \nabla\delta+\delta\nabla=0\,.
\end{equation}
The main property of the local vector field $\nabla$ is that it
respects the odd Poisson bracket, i.e.,
\begin{equation}\label{}
    \nabla(f,g)=-(\nabla f,g)+(-1)^{\epsilon(f)+1}(f,\nabla g)\,,
\end{equation}
for any $f,g\in \mathcal{F}|_{U}$.  Squaring $\nabla$, we get an
internal derivation of the odd Poisson algebra:
\begin{equation}\label{n2}
    \nabla^2 f=(R,  f)\,,
\end{equation}
where one can thought of the odd function  $R=c^\gamma c^\beta
c^\alpha f_{\alpha\beta\gamma}^a\pi_a\in \mathcal{F}_{3,1}$  as
the  curvature of $\nabla$. Like a curvature, $R$ obeys the
Bianchi identity
\begin{equation}\label{}
    \nabla R=0\,.
\end{equation}

Now we can extend $\nabla$ from a local coordinate chart $U$ to
the whole $ \mathcal{N}$. To this end, we choose a trivializing
covering $\{\mathcal{U}{}_i\}$ of $M$ together with local
connections $\nabla_i$ on $(E_0\oplus E_1)|_{\mathcal{U}{}_i}$. It
follows from (\ref{xxx}) that on each nonempty intersection
$\mathcal{U}_i\cap \mathcal{U}_i$
\begin{equation}\label{cocycle}
    \nabla_i-\nabla_j=(\delta\phi_{ij},\,\cdot\,)\,,
\end{equation}
for some $\phi_{ij}\in \mathcal{F}_{1,2}|_{\mathcal{U}_i\cap
\mathcal{U}_i}$. Then, on each nonempty intersection
$\mathcal{U}_i\cap \mathcal{U}_j\cap\mathcal{U}_k$ the functions
$\phi_{ij}$ satisfy the relation
\begin{equation}\label{}
\delta(\phi_{ij}+\phi_{jk}+\phi_{ki})=0\,.
\end{equation}
Since the $\delta$-cohomology is trivial on $\mathcal{F}_{1,2}$ we
conclude that
\begin{equation}\label{lll}
\phi_{ij}+\phi_{jk}+\phi_{ki}=\delta \psi_{ijk}\;\;\;\;\;\;\;\;\;
on\;\;\;\mathcal{U}_i\cap \mathcal{U}_j\cap\mathcal{U}_k\,,
\end{equation}
for some $\chi_{ijk}\in \mathcal{F}_{0,3}|_{\mathcal{U}_i\cap
\mathcal{U}_i\cap\mathcal{U}_k}$. Again, from the last equation it
follows that on each nonempty intersection $\mathcal{U}_i\cap
\mathcal{U}_i\cap\mathcal{U}_k\cap\mathcal{U}_l$ of four domains
one has
\begin{equation}\label{four}
\chi_{ijk}-\chi_{jkl}+\chi_{kli}-\chi_{lij}=0\,.
\end{equation}
Notice that Eq. (\ref{cocycle}) does not define $\phi_{ij}$
uniquely as we are free to add to them any $\delta$-closed terms
$\delta \psi_{ij} \in \mathcal{F}_{1,2}$. This modifies the r.h.s.
of Eq. (\ref{lll}) as follows:
\begin{equation}\label{nnn}
\chi_{ijk}\longrightarrow
\chi_{ijk}-(\psi_{ij}+\psi_{jk}+\psi_{ki})\,.
\end{equation}
Eqs. (\ref{four}), (\ref{nnn}) imply that to any collection of
local connections $\nabla_i$  we have associated an element $\chi$
of the second \v Cech cohomology group  with coefficients in
$\mathcal{F}_{0,3}$. Since this group is clearly isomorphic to the
second De Rham's cohomology group of $M$ we can think of $\chi$ as
an element of $H^2(M)$.

Given a partition of unity $\{h^i\}$ subordinated to the covering
$\{\mathcal{U}_i\}$, we  set
\begin{equation}\label{}
\phi_i=\phi_{ij} h^j\,.
\end{equation}
It is not hard to check, using  Rel. (\ref{lll}), that the  new
local connections $\nabla'_i=\nabla_i-(\delta \phi_i,\,\cdot\,)$
already coincide on each intersection
$\mathcal{U}_i\cap\mathcal{U}_j\neq \emptyset$. Thus, there are no
topological obstructions to introducing a partial connection of
the form (\ref{oddvf}) and we can regard $\chi\in H^2(M)$ as an
invariant of $\nabla$.

Combining now the action of $\nabla$ with internal
differentiations of the odd Poisson algebra $\mathcal{F}$, one can
construct a more general connection
$D:\mathcal{F}_{n,\bullet}\rightarrow \mathcal{F}_{n+1,\bullet}$:
\begin{equation}\label{odvf}
   Da=\delta a +\nabla a -(W,a)=\nabla a +(\omega-W, a) \,,
\end{equation}
$W$ being an element of $\mathcal{F}_{2,\bullet}$. if $D^2=0$, we
refer to $D$ as an abelian connection. The condition of $D$ to be
an abelian connection is equivalent to the following equation:
\begin{equation}\label{WW}
\delta W =R+\nabla W+\frac12(W,W)\,.
\end{equation}
The existence of an abelian connection follows from the solubility
of (\ref{WW}). Indeed, substituting expansion
\begin{equation}\label{}
    W=\sum_{k=2}^\infty
    W_k\,,\qquad W_k\in\mathcal{F}_{2,k}\,,
\end{equation}
into  (\ref{WW}) one gets
\begin{equation}\label{}
\begin{array}{l}
\delta W_2= R\,,\\[3mm]
\displaystyle \delta W_{n+1}=\nabla W_{n} +\frac12\sum_{k=2}^{n}
(W_{n-k+2},W_k)\,,\qquad n\geq 2\,.
\end{array}
\end{equation}
Since the $\delta$-cohomology is trivial when evaluated on
$\mathcal{F}_{2, \,k}$ with $k>0$, the first equation is soluble
provided $\delta R=0$. But the last condition immediately follows
from the identities $0=\nabla^2\omega =(R,\omega)=-\delta R$.
Proceeding by induction, one can see that the r.h.s. of the
$(n+1)$-th equation   is $\delta$-closed (and thus is
$\delta$-exact) provided all the previous equations for
$W_2,...,W_n$ have been satisfied.

The main results of this section can be summarized as follows.

\vspace{5mm}

\noindent \textbf{Proposition 5.1}.
 {\it Suppose we are given by the following data:}
\begin{enumerate}
\item {\it a short exact sequence}
$$
  0\rightarrow \mathcal{E}_2\stackrel{d}{\rightarrow}
\mathcal{E}_1\stackrel{X}{\rightarrow}
    {F}\rightarrow 0\,,
$$
{\it where $\mathcal{E}_1\rightarrow M$, $\mathcal{E}_2\rightarrow
M$ are vector bundles over a smooth manifold $M$, ${F}$ is an
integrable subbundle of the tangent bundle $TM$, and $d$, $X$ are
$M$-morphisms of the vector bundles (not necessarily of constant
rank);}

\item {\it a non-degenerate, skew-symmetric, bilinear form
$\omega$ on $\mathcal{E}_1$ inducing a Poisson bi-vector field on
the base manifold:
$$
    \alpha = \langle\omega , X\wedge X\rangle \in
    \wedge^2TM\,,\;\;\;\;\;\;\;\;
[\alpha,\alpha]=0\,.$$ (Here we identify $X :
\mathcal{E}_1\rightarrow {F}\subset TM$ with a section of
$\mathcal{E}^\ast_1\otimes TM$.)}
\end{enumerate}
{\it Then to each set of such data one can associate

\begin{enumerate}
    \item an invariant $\chi$ taking value in the second group of
    De Rham's cohomology of $M$, and
    \item
a bundle of odd Poisson algebras $\mathcal{F}$ over $M$ together
with an abelian connection $D$ differentiating $\mathcal{F}$ such
that the condition $D^2=0$ generates all structure relations
arising from the integrability of $F$ and the Jacobi identity for
$\alpha$.
\end{enumerate}
}

The generating procedure stated above could be viewed as starting
point for quantizing general factorizable brackets associated with
symplectic $2$-algebroids along the lines of Sections 3, 4.


\begin{thebibliography}{10}

\bibitem{Berezin}  F.A. Berezin, \textit{``Quantization''}, Izv. Akad. Nauk. {\bf 38} (1974) 1116-1175;
\textit{``General concept of quantization''}, Comm. Math. Phys.
{\bf 40} (1975) 153-174.

\bibitem{BFFLS}  F. Bayen, M. Flato, C. Fronsdal, A. Lichnerowicz, and
D. Sternheimer, {\it ``Deformation Theory and Quantization. 1.
Deformation of Symplectic Structures"}, Ann. Phys.(N.Y.) {\bf 111}
(1978), 61.

\bibitem{Fedosov} B.V. Fedosov, {\it ``A Simple Geometrical Construction of
Deformation Quantization"}, J. Diff. Geom. {\bf 40} (1994) 213.

\bibitem{Fedosovbook} B.V. Fedosov, Deformation Quantization and Index
Theory, Berlin, Germany: Akademie-Verl. (1996) (Mathematical
Topics: 9).

\bibitem{Kontsevich} M. Kontsevich, {\it ``Deformation quantization of
Poisson manifolds, I"}, Lett. Math. Phys. \textbf{66} (2003) 157.

 \bibitem{CFT} A. S. Cattaneo, G. Felder, L. Tomassini, {\it
 ``From local to global deformation quantization of Poisson
manifolds"}, Duke Math. J. \textbf{115} (2002) 329.

\bibitem{Dolg} V. Dolgushev, {\it ``Covariant and Equivariant Formality
Theorem"}, Adv. Math. \textbf{191} (2005) 147-177.

\bibitem{CF} A.S. Cattaneo and G. Felder, {\it ``A path integral approach
to the Kontsevich quantization formula"}, Commun. Math. Phys.
\textbf{212} (2000) 591.


\bibitem{BV1} I.A. Batalin and G.A. Vilkovisky, {\it ``Gauge algebra
and quantization''}, Phys. Lett. \textbf{B102} (1981) 27-31.

\bibitem{AK} A. Alekseev and Y. Kosmann-Schwarzbach, {\it ``Manin pairs and moment
maps''}, math.DG/9909176.

\bibitem{NT} R. Nest and  B. Tsygan, \textit{``Deformations of symplectic Lie
algebroids, deformations of holomorphic symplectic structures, and
index theorems''}, Asian J. Math. \textbf{5} (2001) 599-635.

\bibitem{Vaisman} I. Vaisman, {\it ``Fedosov Quantization on Symplectic Ringed
Spaces"}, math.SG/0106070.


\bibitem{Rieffel} M. A. Rieffel, {\it ``Deformation Quantization for Action of
$R^d$"}, Memoirs A.M.S. {\bf 506}, AMS, Providence, 1993.

\bibitem{Xu}   P. Xu, {\it ``Triangular dynamical r-matrices and quantization"},
Adv. Math. \textbf{166}, 1 (2002) 1-49.

\bibitem{DILSh} V.A. Dolgushev, A.P. Isaev, S.L. Lyakhovich and
A.A. Sharapov, {\it ``On the Fedosov deformation quantization
beyond the regular Poisson manifolds"}, Nucl.Phys. {\bf B645}
(2002) 457-476.


\bibitem{BFV} E.S. Fradkin  and G.A. Vilkovisky, {\it ``Quantization
of relativistic systems with  constraints'',} Phys. Lett. {\bf
B55} (1975) 224-226; \, I.A. Batalin  and  G.A. Vilkovisky, {\it
``Relativistic S-matrix of dynamical systems with boson and
fermion constraints'',} Phys. Lett. {\bf B69} (1977) 309-312.

\bibitem{BFV2} I.A. Batalin  and E.S. Fradkin,
{\it  ``A generalized canonical formalism and quantization of
reducible gauge theories'',} Phys.Lett. {\bf B122 }(1983) 157-164.


\bibitem{HT} M. Henneaux and C. Teitelboim, Quantization of Gauge
Systems (Princeton U.P., NJ, 1992).


\bibitem{GL} M.A. Grigoriev and S.L. Lyakhovich, {\it ``Fedosov
Deformation Quantization as a BRST Theory"}, Commun. Math. Phys.
{\bf 218} (2001) 437.

\bibitem{BGL} I.A. Batalin, M.A. Grigoriev, S.L. Lyakhovich,
{\it ``Star Product for Second Class Constraint Systems from a
BRST Theory"}, Theor. Math. Phys. {\bf 128} (2001) 1109-1139;
hep-th/0101089.



\bibitem{Vain} A. Vaintrob, {\it ``Lie algebroids and homological vector fields''
}, Uspekhi Mat. Nauk, {\bf 52} (1997) 428-429.

\bibitem{Sev} P. \v{S}evera, {\it ``Some title containing the words ``homotopy'' and
``symplectic'', e.g. this one''}, math.SG/0105080.

\bibitem{Vor} Th. Th. Voronov, {\it ``Graded manifolds and Drinfeld doubles for Lie
bialgebroids''}, Contemp. Math. {\bf 315} (2002) 131-168.

\bibitem{Roy} D. Roytenberg, {\it ``On the structure of graded symplectic supermanifolds
and Courant algebroids''}, Contemp. Math. {\bf 315} (2002)
169-185.

\bibitem{WCdS} A. Cannas da Silva and A. Weinstein, Geometric
Models for Noncommutative Algebras, \textit{Berkeley Mathematics
Lecture Notes}, vol.\textbf{10 } (AMS, Providence, RI, 1999).

\bibitem{Sussman} H. Sussman, {\it ``Orbits of families of vector fields and integrability
of distributions"}, Trans. Amer. Math. Soc. {\bf 180}, 171-188.

 \bibitem{MX} K.C.H. Mackenzie and P. Xu, {\it ``Lie bialgebroids and Poisson
 groupoids"}, Duke Math. J. {\bf 73} (1994) 415-452.

\bibitem{El}  A.G. Elashvili, {\it ``Frobenius Lie algebras"}, Funct. Anal.
Applic. {\bf 16} (1982) 94-95.

\bibitem{BT}I.~A.~Batalin and I.~V.~Tyutin, {\it ``Existence Theorem For The
Effective Gauge Algebra In The Generalized Canonical Formalism
With Abelian Conversion Of Second Class Constraints"}, Int.J. Mod.
Phys. A {\bf 6} (1991) 3255.

\bibitem{Courant} T. Courant, {\it Dirac manifolds}, Trans. Amer.
Math. Soc. {\bf 319} (1990) 631-661.

\bibitem{EW} C. Emmrich and A. Weinstein, {\it ``The differential geometry of Fedosov's
quantization"}, Lie Theory and Geometry, 217-239, Progr. Math.
{\bf 123}, Birkh\"auser Boston, Boston, 1994.


\end{thebibliography}
\end{document}